\documentclass[ reprint,superscriptaddress, amsmath,amssymb, aps, longbibliography]{revtex4-1}

\usepackage[colorlinks=true,linkcolor=blue,urlcolor=blue,citecolor=blue]{hyperref}
\usepackage{overpic}
\usepackage{upgreek}
\usepackage{physics}
\usepackage{graphicx}
\usepackage{dcolumn}
\usepackage{bm}
\usepackage{outlines}
\usepackage[dvipsnames]{xcolor}
\usepackage{gensymb}
\usepackage{subcaption}
\usepackage{caption}
\usepackage{tikz}  
\usepackage{comment}
\usepackage{booktabs} 
\usepackage{xcolor}

\usepackage{pdfpages}
\makeatletter
\AtBeginDocument{\let\LS@rot\@undefined}
\makeatother

\newcommand{\mitll}{MIT Lincoln Laboratory, Lexington, MA 02421, USA}
\newcommand{\mitcba}{Center for Bits and Atoms, Massachusetts Institute of Technology, Cambridge, MA 02139, USA}

\begin{document}
\title{Elimination of Flux Trapping in Superconducting Circuits in Ambient Magnetic Fields}

\author{Rohan T.~Kapur}
\altaffiliation{Present Address: Department of Physics, Harvard University, Cambridge, MA 02138, USA}
\email{rohankapur@fas.harvard.edu}
\affiliation{\mitll}
\author{Alex Wynn}
\email{alexander.wynn@cba.mit.edu}
\affiliation{\mitll}
\affiliation{\mitcba}
\author{Sergey K.~Tolpygo}
\email{sergey.tolpygo@ll.mit.edu}
\author{Neel Parmar}
\author{Anil Mankame}
\author{Adam A.~Libson}
\author{Rabindra Das}
\author{Michele Kelley}
\author{Pauli Kehayias}
\author{Nathaniel J.~O'Connor}
\author{Collin N.~Muniz}
\author{Justin L.~Mallek}
\author{Jennifer M.~Schloss}
\affiliation{\mitll}


\begin{abstract}
Superconductor digital electronics and quantum computing with superconducting qubits are promising next-generation computing technologies. When cooled down or operated in the presence of a nonzero background magnetic field $B_r$, superconducting thin films comprising the circuits can trap magnetic vortices that can degrade circuit or qubit performance. In this work, we report a practical solution for eliminating flux trapped during cooldown in ambient magnetic fields, $B_r\leq 60$ $\upmu$T, based on controlled local thermal gradients and moats, etched holes in the superconducting films of the circuit. Thermal gradients created by integrated on-chip resistive heaters move vortices towards the moats, where they become trapped away from circuitry regions and pinning sites. Using magnetic imaging and electrical circuit readout, we demonstrate that this approach is capable of removing magnetic flux trapped during field cooling and magnetic flux nucleated by circuit operation. If used in an environment with basic magnetic shielding, this solution is capable of suppressing all magnetic flux in a large-scale circuit, overcoming one of the long-standing challenges preventing high-performance scalable computing using superconductors.
\end{abstract} 
\date{\today}

\maketitle
\section{Introduction}
Superconducting circuits are attractive platforms for next-generation classical and quantum computing. Superconductor digital electronics have demonstrated 100$\times$ improvements in energy efficiency and clock speeds relative to complementary metal oxide (CMOS) electronics~\cite{SCEreviewVanDuzer, SCEreviewSpringer, SCEreviewIEEE2024}, while superconducting qubits have emerged as a leading platform for scalable quantum computing~\cite{nakamura1999coherent,koch2007charge,barends2013coherent,krantz2019quantum,manucharyan2009fluxonium,devoret2013superconducting}. Despite these advances, both technologies remain far from practical large-scale deployment. Scaling is limited by conventional barriers, such as fabrication technology, control and readout complexity, circuit cross talk, and on-chip power distribution, and superconducting phenomena, particularly magnetic flux trapping \cite{shregOld, nagasawa_sce_scaling, ac_power_sfq, Herr_2015,tolpygo2022scalability, Mutsuo_HIDAKA20212020SUI0002, ayala_monolithic_infra,krantz2019quantum,devoret2013superconducting,megrant2025scaling}.

Superconductor digital circuits typically use thin-film superconductors such as Nb, NbN, and NbTiN  for their large critical current densities, high critical magnetic fields, and fabrication advantages, while superconducting qubits also use thin-film Al and Ta for their low RF losses \cite{likharev2012superconductor, tolpygo2016superconductor, tahara2002superconducting, tolpygo2023progress, nieto2023flexible,pokhrel2024nbtin,krantz2019quantum,devoret2013superconducting,megrant2025scaling,chiaro2016dielectric}. When cooled through the superconducting transition temperature ($T_c$) in the presence of a background magnetic field ($B_r \neq 0$), or exposed to magnetic fields exceeding the lower critical field $B_{c1}$, these materials form a mixed state consisting of quantized magnetic flux (vortices). These vortices can strongly interfere with and degrade circuit and qubit performance  \cite{likharev2012superconductor, tolpygo2016superconductor, tahara2002superconducting, tolpygo2023progress, nieto2023flexible, pokhrel2024nbtin, song2009reducing,nsanzineza2014trapping,chiaro2016dielectric,bafia2025quantifying,krantz2019quantum}.

Solving magnetic flux trapping requires the removal of nearly all vortices in the vicinity of critical circuitry, as a single vortex can disrupt device operation \cite{10842353, shregNew, nsanzineza2014trapping}. To mitigate flux trapping in superconducting circuits, a variety of strategies have been developed. These include minimizing background magnetic fields using multilayer magnetic shielding ($B_r \approx 50$ nT) or superconducting shields with active field shimming ($B_r \approx 1$ nT). On-chip approaches include etching holes in the superconducting film, commonly referred to as moats or antidots, to passively attract and trap flux away from sensitive regions~\cite{semenov2016moats,IBMmoats,ssmMoats1995,fourie2021experimental,colauto2020controlling,kapur2026mitigationmagneticfluxtrapping,song2009reducing} or actively transporting vortices using current-induced Lorentz force or AC-current ratcheting ~\cite{ratchetEffect, ac_deflux_squid, wang2005manipulating}. 

While many mitigation strategies have demonstrated significant flux removal, a solution capable of fully suppressing vortex formation has not been realized. For example, the largest superconducting qubit systems demonstrated to date contain several thousands qubits and a few million Josephson junctions on relatively large-area chips~\cite{castelvecchi2023ibm,dwave_advantage2_4400q_2025}. Even at $B_r=$ 1 nT, a 2-$\mathrm{cm}^2$ superconducting film can trap a hundred of vortices, since the number of trapped flux quanta in bare films scales as $N_{\Phi_0} = B_rA/\Phi_0$, where $\Phi_0 \approx 2.07 \times 10^{-15}~\mathrm{Wb}$ is the magnetic flux quantum, and $A$ is the superconducting film area. As superconducting circuits scale further, the number of trapped vortices may increase owing to both the larger superconducting area and the increased difficulty of maintaining ultralow residual magnetic fields over the entire circuit area. The problem is further exacerbated by film  defects and circuit patterns, which introduce pinning sites and enhance vortex trapping even in the presence of mitigation strategies such as moats~\cite{kapur2026mitigationmagneticfluxtrapping}. Furthermore, magnetic fields generated by DC and AC operating currents can also generate additional vortices.

In this work, we present a magnetic flux trapping suppression strategy based on locally generated thermal gradients and moats, which can eliminate trapped magnetic flux in ambient background fields $B_r \lesssim 60~\upmu\mathrm{T}$. We demonstrate the approach using a simple Josephson junction-based superconducting circuit, with the underlying mechanism broadly generalizable to any superconducting circuit. Thermal gradients generated by on-chip resistive heaters drive vortices toward higher-temperature regions containing aligned moats and simultaneously reduce the strength of local pinning sites ~\cite{geng1992sweeping, veshchunov2016optical}. The moats then trap the vortices away from active circuitry, where they remain even after the heaters are turned off. The effectiveness of the thermal gradients is characterized using widefield magnetic field microscopy using nitrogen-vacancy (NV) centers in diamond \cite{qswift_apparatus_paper} and concurrent electrical measurements of Josephson junction (JJ) chains serving as model circuitry. We observe significant changes in the magnetic flux distribution and electrical performance of the JJ chains as a function of heater power, which are consistent with the transport of magnetic flux from near the JJ chains to the moats. While requiring space in a single layer of the circuit ($25\% \text{ in the design used, potentially reducible to} \leq 5\%$) and independent power supply to operate, this approach is likely capable of suppressing trapped magnetic flux in large-scale superconducting circuits operating with basic magnetic shielding.
\section{Methods}
\begin{figure*}[htbp]
    \centering
    
    \begin{minipage}[t]{0.38\textwidth}
        \centering
        \begin{overpic}[width=\linewidth, trim=190pt 0pt 180pt 0pt, clip]{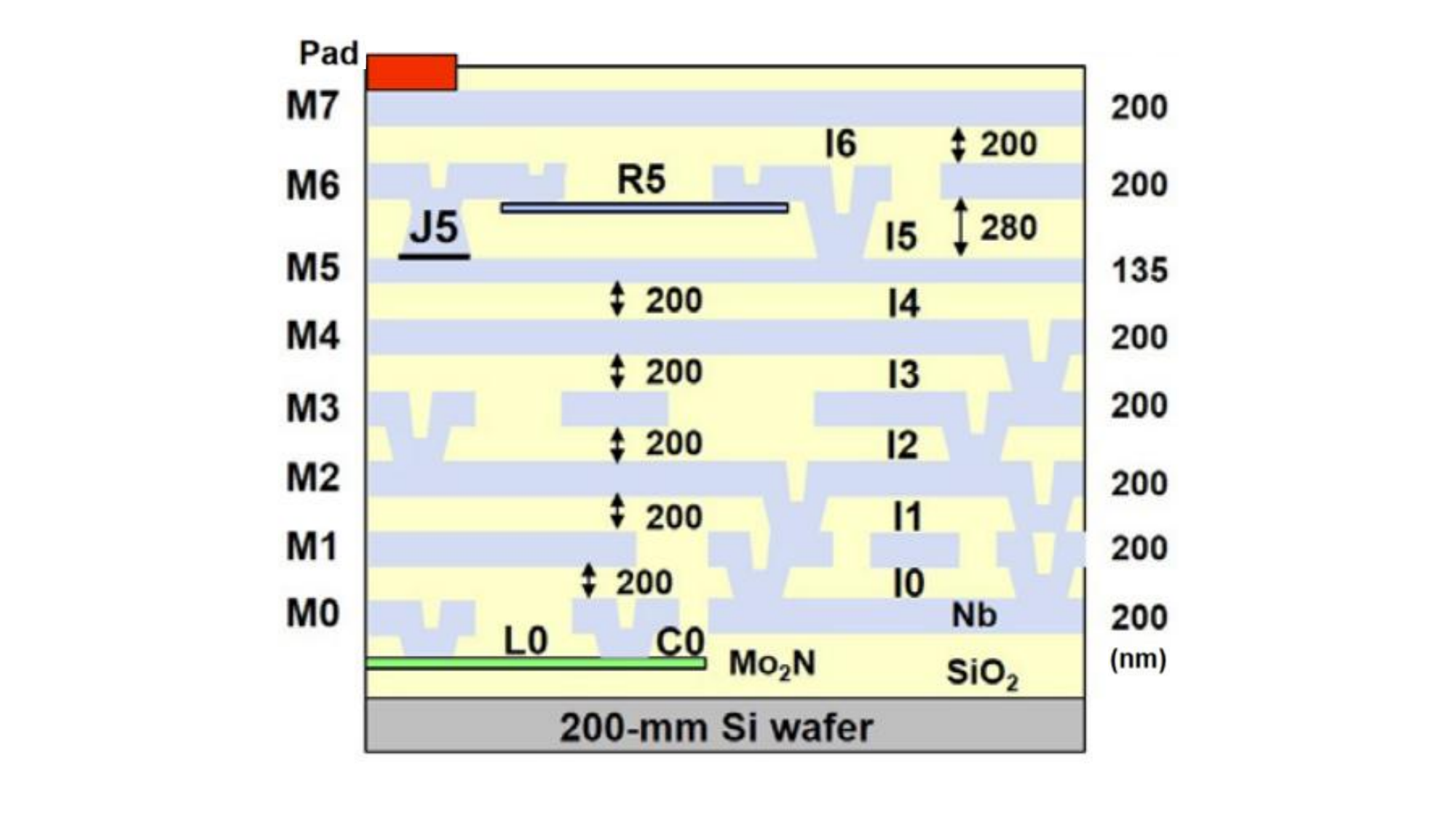}
            \put(-8,83){\footnotesize\textbf{(a)}}
        \end{overpic}
    \end{minipage}
    \hfill
    \begin{minipage}[t]{0.6\textwidth}
        \centering
        \begin{overpic}[width=\linewidth, trim=0pt 0pt 0pt 0pt, clip]{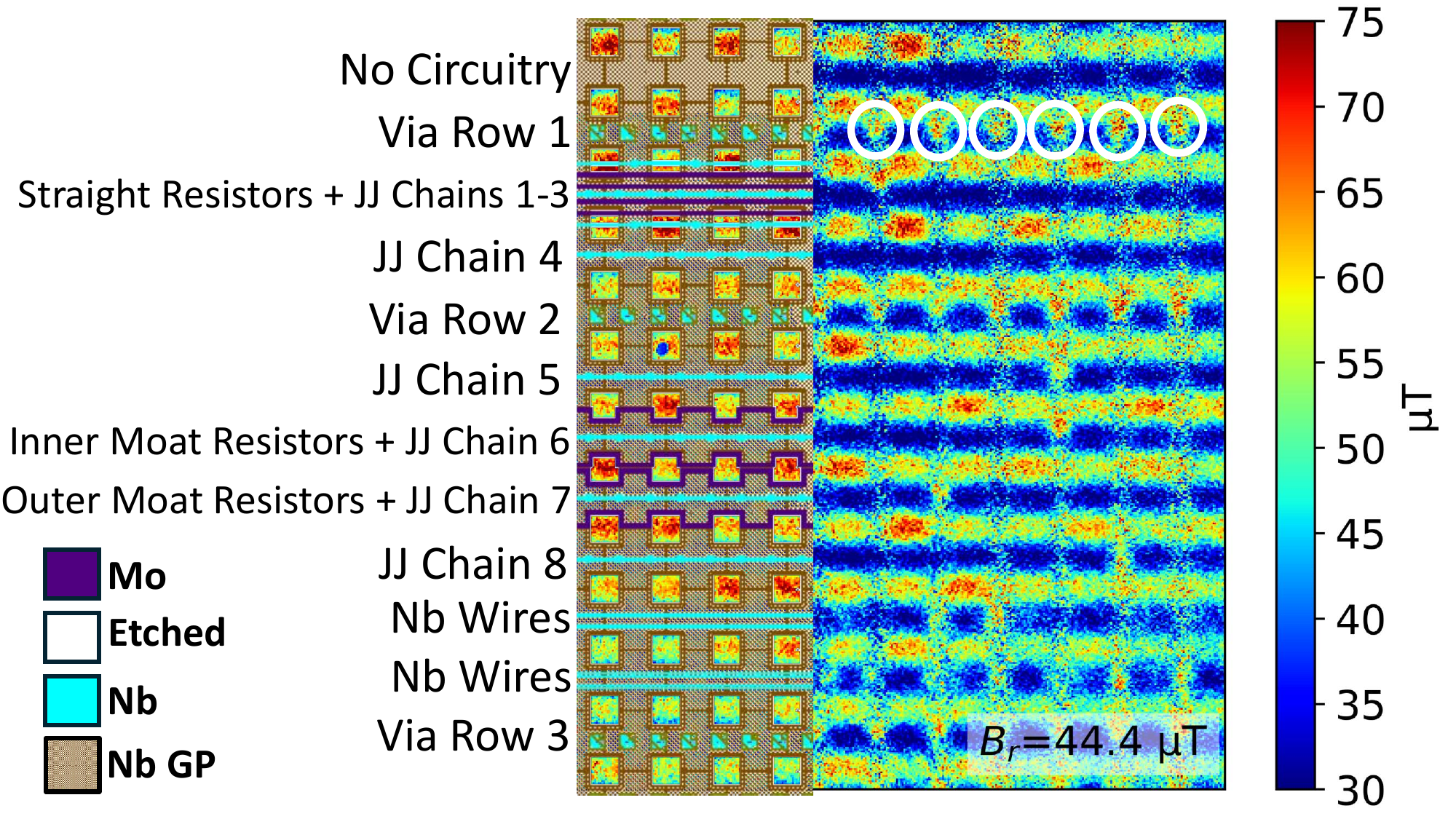}
            \put(-3,53){\footnotesize\textbf{(b)}}
        \end{overpic}
    \end{minipage}

    \caption{(a) Schematic cross section of the MIT LL SFQ5ee process, showing superconducting Nb (M0,...,M7) and $\mathrm{SiO_2}$ dielectric layers \cite{SFQ5ee}. The chip used in this work has three Nb ground plane layers (M1, M4, and M7) patterned with congruent \(10~\upmu\mathrm{m}\times10~\upmu\mathrm{m}\) moats spaced by \(10~\upmu\mathrm{m}\). The circuitry involves resistor layer, R5, and Josephson junctions, J5, interconnected by wires in layers M5 and M6. All metal layers except the ground planes contain also Nb fill structures -- isolated \(5~\upmu\mathrm{m}\times5~\upmu\mathrm{m}\) Nb squares placed with \(10~\upmu\mathrm{m}\times20~\upmu\mathrm{m}\) pitch. (b) Layout of circuitry overlaid with a magnetic image for \(B_r=44.4~\upmu\mathrm{T}\) and no current applied to the circuitry. Example vortices are circled in white. We observe differences in flux trapping behavior across the chip as a function of local circuitry. Isolated vortices appear for \(B_r\leq30~\upmu\mathrm{T}\), primarily near vias, with the first vortices forming in the film at \(B_r=25.2~\upmu\mathrm{T}\). For \(B_r\geq30~\upmu\mathrm{T}\), vortices form throughout the film, with enhanced trapping near vias and Nb wires and occasional trapping near JJ chains and Mo resistors.}
    \label{fig:passive_gds}
\end{figure*}

We designed and fabricated a superconducting electronic circuit to evaluate the effectiveness of thermal gradients in mitigating trapped magnetic flux. The chip was based on the MIT Lincoln Laboratory SFQ5ee process \cite{SFQ5ee} and incorporated three Nb ground planes in the M1, M4, and M7 layers (Fig.~\ref{fig:passive_gds}a). These ground planes were patterned with square moat arrays (10~$\upmu$m side length, 10~$\upmu$m spacing) and integrated with JJ chains placed between the moats and serving as model circuitry (Fig.~\ref{fig:passive_gds}b). All other metal layers in the circuit were patterned with Nb fill structures, which consisted of $5$$~\upmu\mathrm{m}$ $\times~5$ $\upmu\mathrm{m}$ squares spaced by $5$ $\upmu$m in the horizontal direction and $15$ $\upmu$m in the vertical direction.

JJ chains consisting of 300 individual Josephson junctions connected in series were chosen as the model circuitry due to their fabrication simplicity, reliability, and ability to function as local thermometers. Thermal gradients were generated using molybdenum (Mo) resistors fabricated in the standard SFQ5ee resistor layer, R5 (Fig.~\ref{fig:passive_gds}a). These resistors are 40~nm thick and 1~$\upmu$m wide, with a sheet resistance of approximately 2~$\Omega/\square$. Their relatively high resistance enables the generation of a large localized heat power at low applied currents, which minimizes magnetic fields and enables scaling to chip-wide heater implementations (Appendix~\ref{subsec:chip_scale_circuit_scaling}).

 Using magnetic imaging, we first examined if the presence of resistors affects flux trapping, comparing the circuit regions with and without resistive components. We then used current applied to the resistors to generate thermal gradients and studied their effect on flux distribution in superconducting films in two unique heater configurations: (i) \emph{outer moat resistors}, consisting of two independent resistors aligned with the outer edges of adjacent moat rows, and (ii) \emph{inner moat resistors}, consisting of two independent resistors aligned with the inner edges of adjacent moat rows. In both designs, the resistors span the width of the chip, producing higher-temperature regions along the resistors with temperature gradient towards the resistors. These gradients are intended to drive flux toward the heated moat regions and away from the JJ chains between the moats.

To characterize the effect of thermal gradients on flux trapping, we magnetically imaged the circuit in field-cooled experiments at various $B_r$ and currents applied to the resistors. Magnetic field imaging was conducted using a cryogenic widefield NV-diamond microscope, which provides quantitative maps of the out-of-plane magnetic field across the device under test (DUT) \cite{qswift_apparatus_paper}. The microscope offers rapid magnetic field imaging of vortices across millimeter-scale areas with micron-scale spatial resolution. In addition, the microscope setup enables the application and measurement of DC and low-frequency AC signals to the DUT, which we use to generate thermal gradients and electrically probe the JJ chains (Appendix~\ref{subsec:microscope_active_upgrade}). We further validated the effectiveness of the thermal gradients in removing trapped flux through current-voltage (IV) measurements of the JJ chains done in a separate liquid helium dunk (LHe) probe.

\section{Results}
\subsection{Effect of Circuitry on Magnetic Flux Trapping}
The test chip contained a variety of circuit elements, including Nb wires, interlayer vias, JJ chains, Nb wires, and Mo resistors, as well as three electrically connected ground planes patterned with congruent moats. We magnetically imaged different regions to investigate the influence of circuitry on magnetic flux trapping in the absence of applied currents. At low magnetic fields ($B_r \leq$ 30 $\upmu$T), we observed isolated vortex formation, with most vortices appearing near vias and the first vortices observed in the film at $B_r=25.2$ $\upmu$T. These vortices tended to appear in the same locations across multiple temperature cycles. The majority of the chip area, however, remains vortex-free in this regime, indicating that most flux is accommodated by the moats and that vias are likely acting as vortex nucleation and/or pinning sites. 

At higher magnetic fields, we observed significant differences in flux trapping behavior across the top superconducting film. Areas of the film above the vias between the lower-laying layers appear to \emph{enhance} vortex density with all measured rows of vias having significantly more flux than rows without vias or with other circuit elements. The presence of Nb wires between the M4 and M7 ground planes also appear to enhance the vortex areal density in the M7 ground plane between the moats, although this effect is less consistent than that of the vias. The JJ chains appeared to weakly enhance vortex density with isolated vortices appearing around JJ chains. However, the vortex density is not consistent across chains, indicating that vortex nucleation in these regions may be related to fabrication defects instead of the properties of the JJ chains themselves. Mo resistors, which were of most concern for these experiments, seem to have a negligible impact on flux trapping with all three imaged resistor regions having just a few more vortices than the bare film region. It may be that these few vortices formed due to the presence of nearby JJ chains because JJ chains without the adjacent resistors were observed to increase the vortex density. However, this requires further investigation. In general, the presence of resistors does not increase the trapped flux density in the film appreciably. An example image of the described flux distribution in various regions of the circuit is shown in Fig.~\ref{fig:passive_gds}b for $B_r=44.4$ $\upmu$T.
\begin{figure}[htbp]
  \centering
  \captionsetup{font=small}

  \begin{overpic}[width=0.95\columnwidth,trim=0pt 0pt 40pt 0pt, clip]{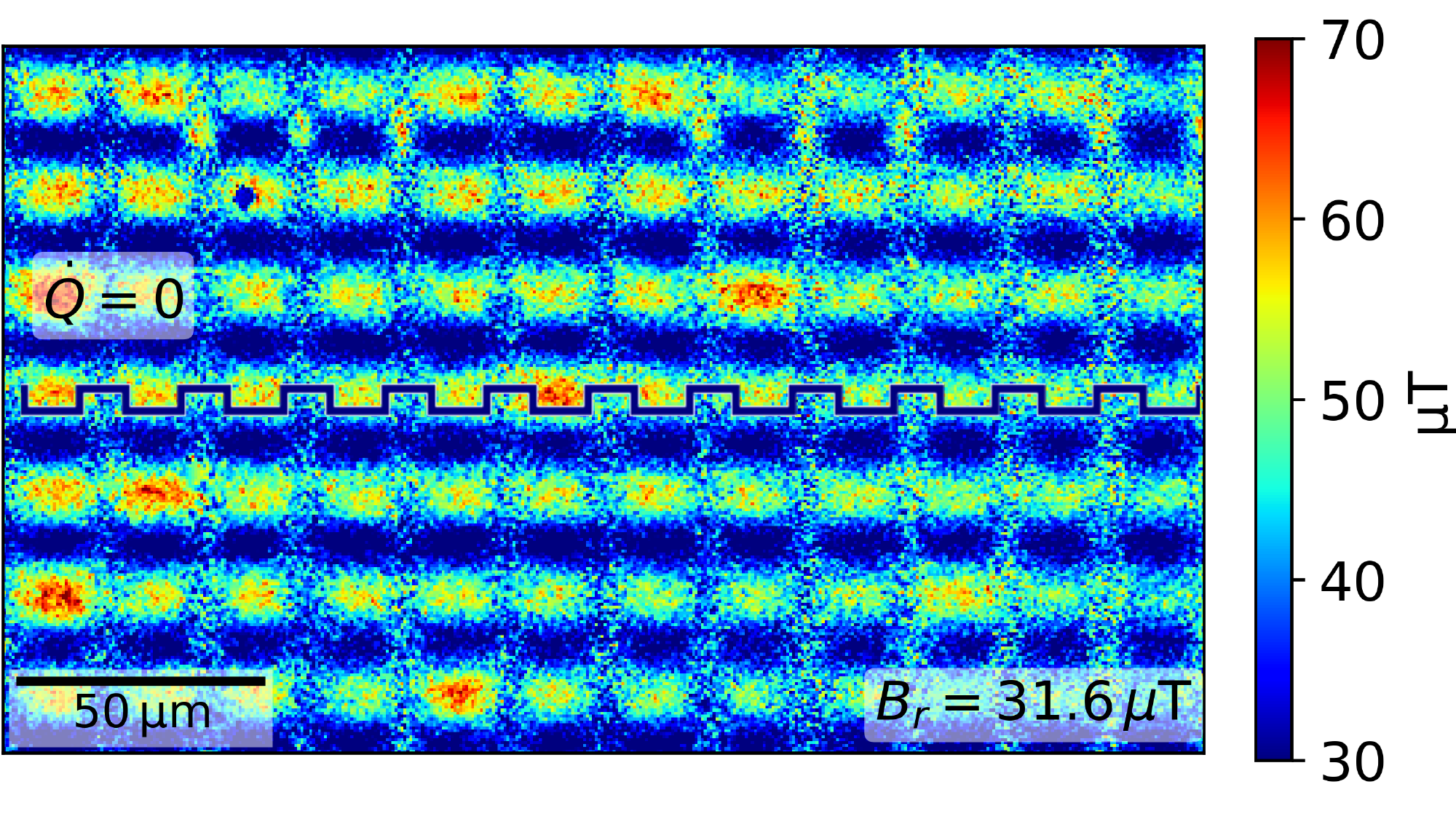}
    \put(-2,57){\footnotesize\textbf{(a)}}
  \end{overpic}

  \vspace{0.5em}

  \begin{overpic}[width=0.95\columnwidth,trim=0pt 0pt 40pt 0pt, clip]{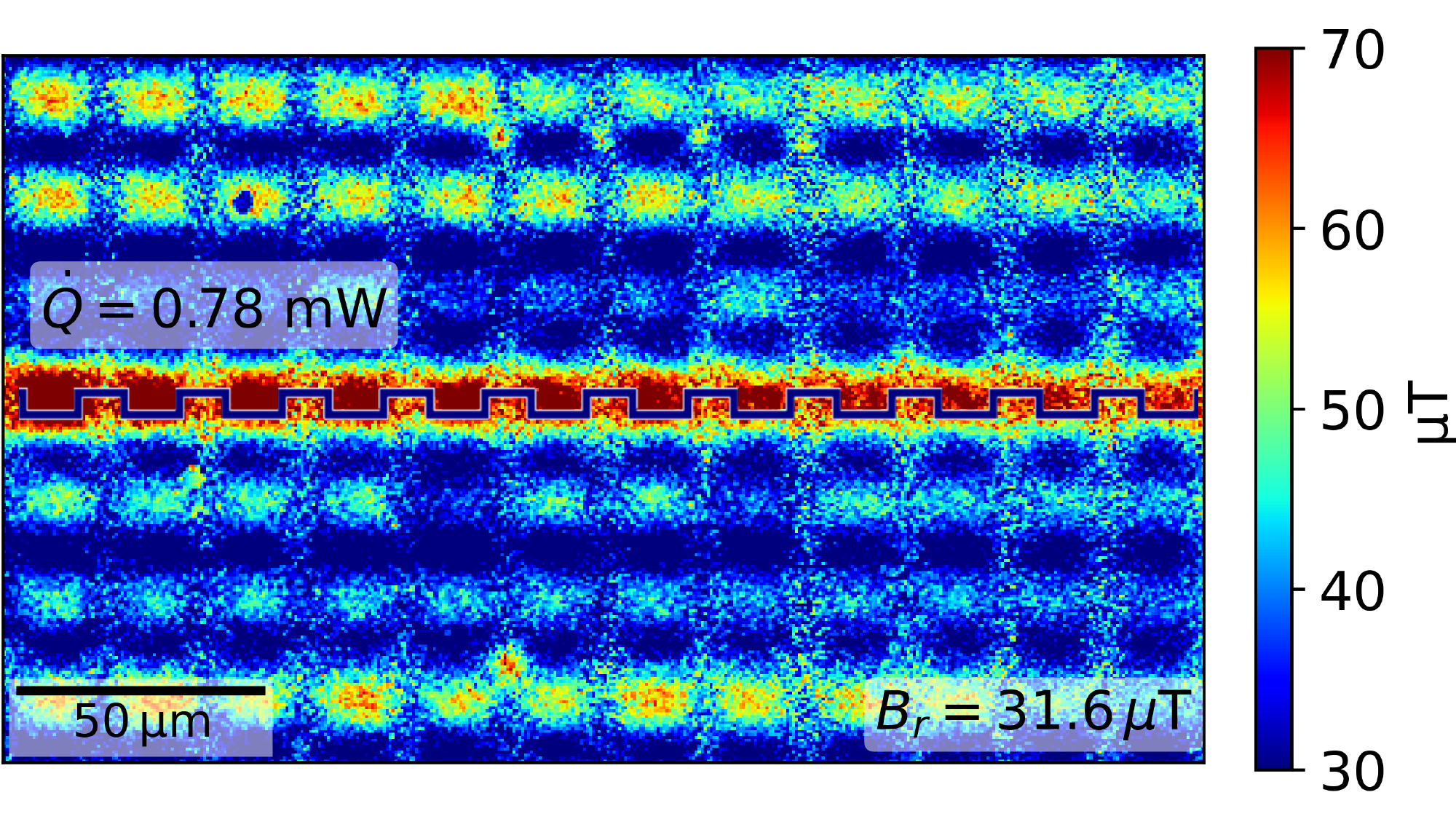}
    \put(-2,57){\footnotesize\textbf{(b)}}
  \end{overpic}

  \vspace{0.5em}

  \begin{overpic}[width=0.95\columnwidth,]{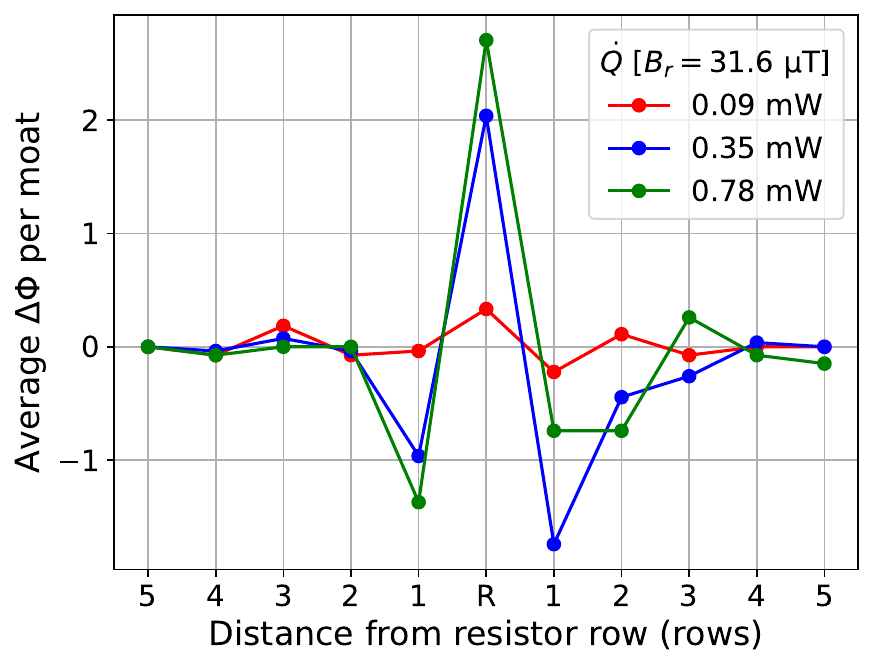}
    \put(-2,75){\footnotesize\textbf{(c)}}
  \end{overpic}

  \caption{(a-b) Magnetic image of the flux distribution for $B_r=31.6$ $\upmu$T and (a) $\dot{Q}$ = 0 mW and (b) $\dot{Q}$ = 0.78 mW. As seen in the image, applying current to the resistor during cooldown significantly changes the flux distribution, with the moats aligned with the resistor containing significantly more flux quanta than adjacent moats. The images and color bars in (a) and (b) are in units of $\upmu$T (c) Change in $\Phi_0$ per moat relative to the $\dot{Q}=0$ case, $\Delta\Phi$, for each moat row (centered about the resistor row) as a function of $\dot{Q}$. }
  \label{fig:single_outer_moat_resistor_measurement}
\end{figure}
\subsection{Flux Expulsion Using a Single Resistive Heater}
\begin{figure*}[htbp]
  \centering
  \captionsetup{font=small}
  
  \begin{overpic}[width=0.48\textwidth, trim=0pt 0pt 50pt 0pt, clip]{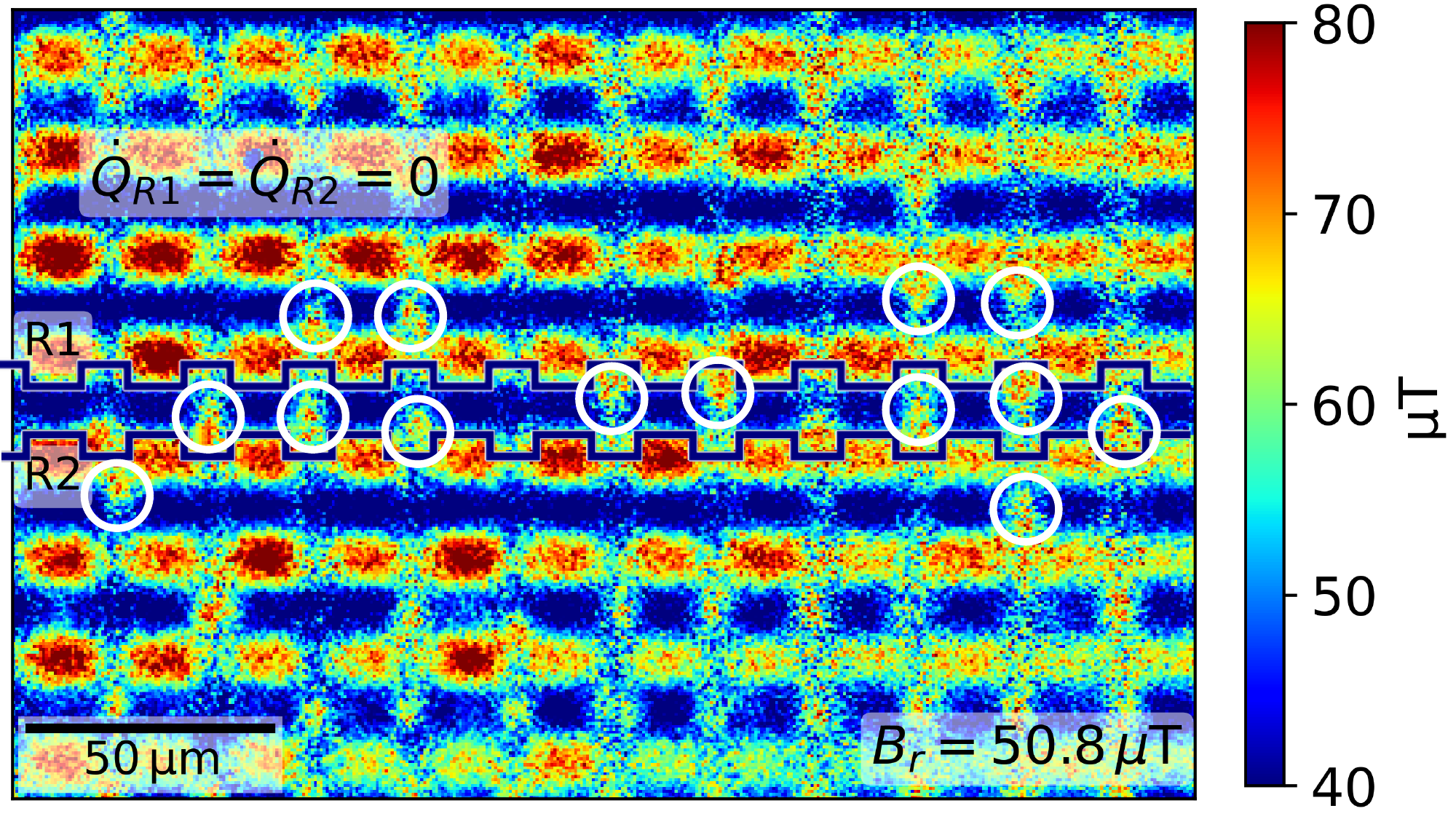}
    \put(-6,52){\footnotesize\textbf{(a)}}
  \end{overpic}
  \hspace{0.02\textwidth}
  \begin{overpic}[width=0.48\textwidth,trim=0pt 0pt 50pt 0pt, clip]{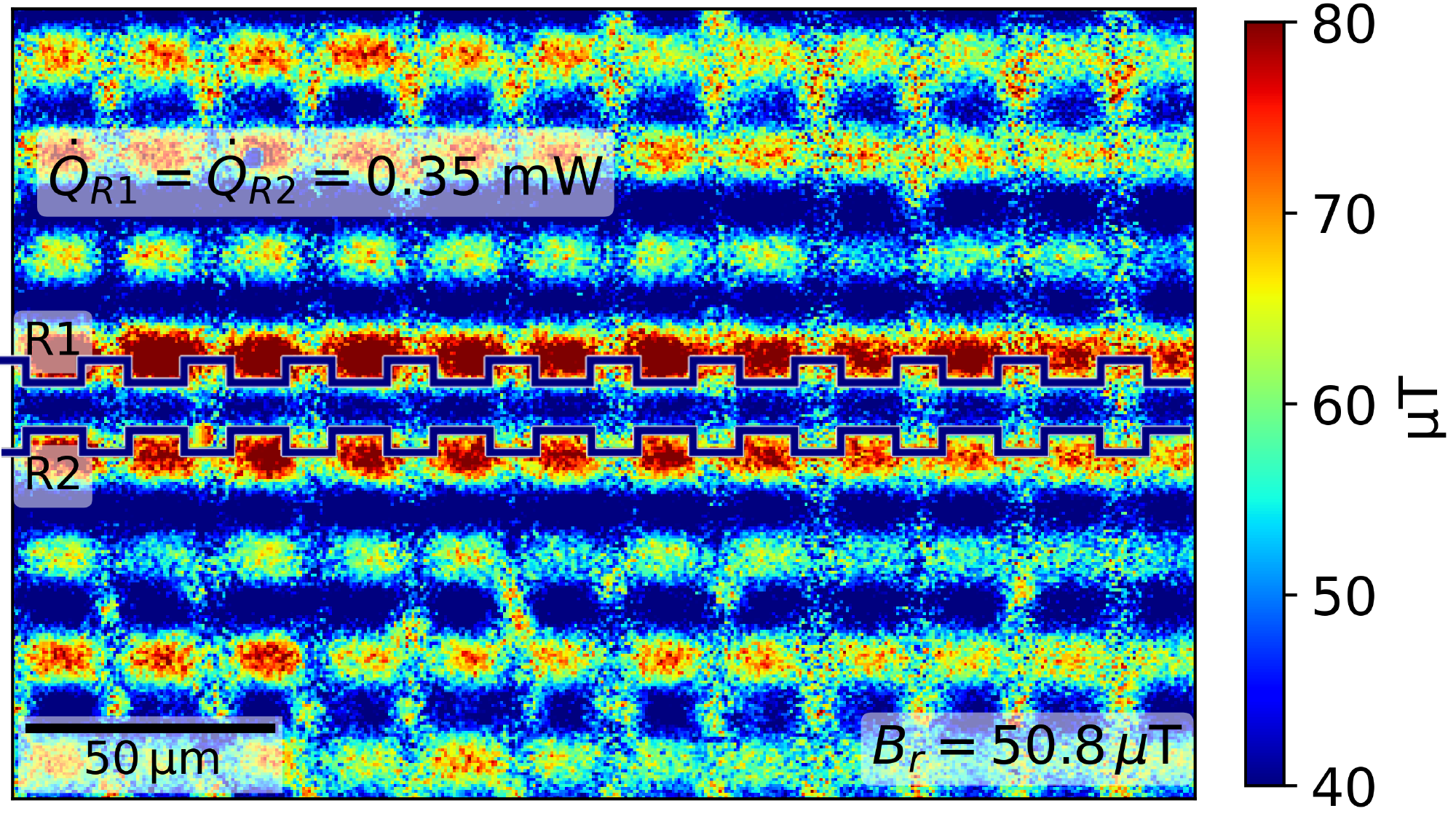}
    \put(-6,52){\footnotesize\textbf{(b)}}
  \end{overpic}

  \vspace{0.75em}

  \begin{overpic}[width=0.48\textwidth,trim=0pt 0pt 40pt 0pt, clip]{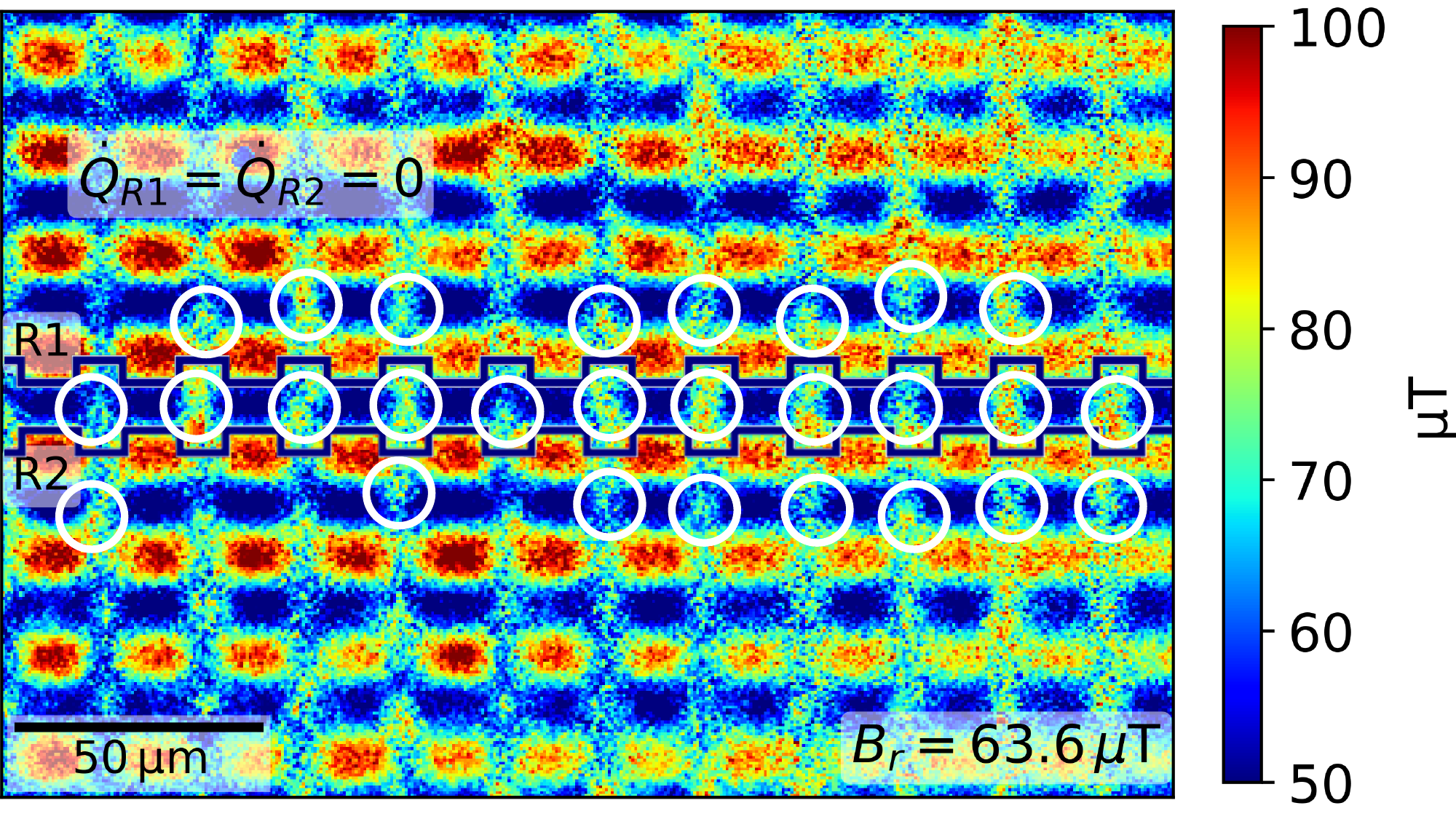}
    \put(-6,52){\footnotesize\textbf{(c)}}
  \end{overpic}
  \hspace{0.02\textwidth}
  \begin{overpic}[width=0.48\textwidth,trim=0pt 0pt 40pt 0pt, clip]{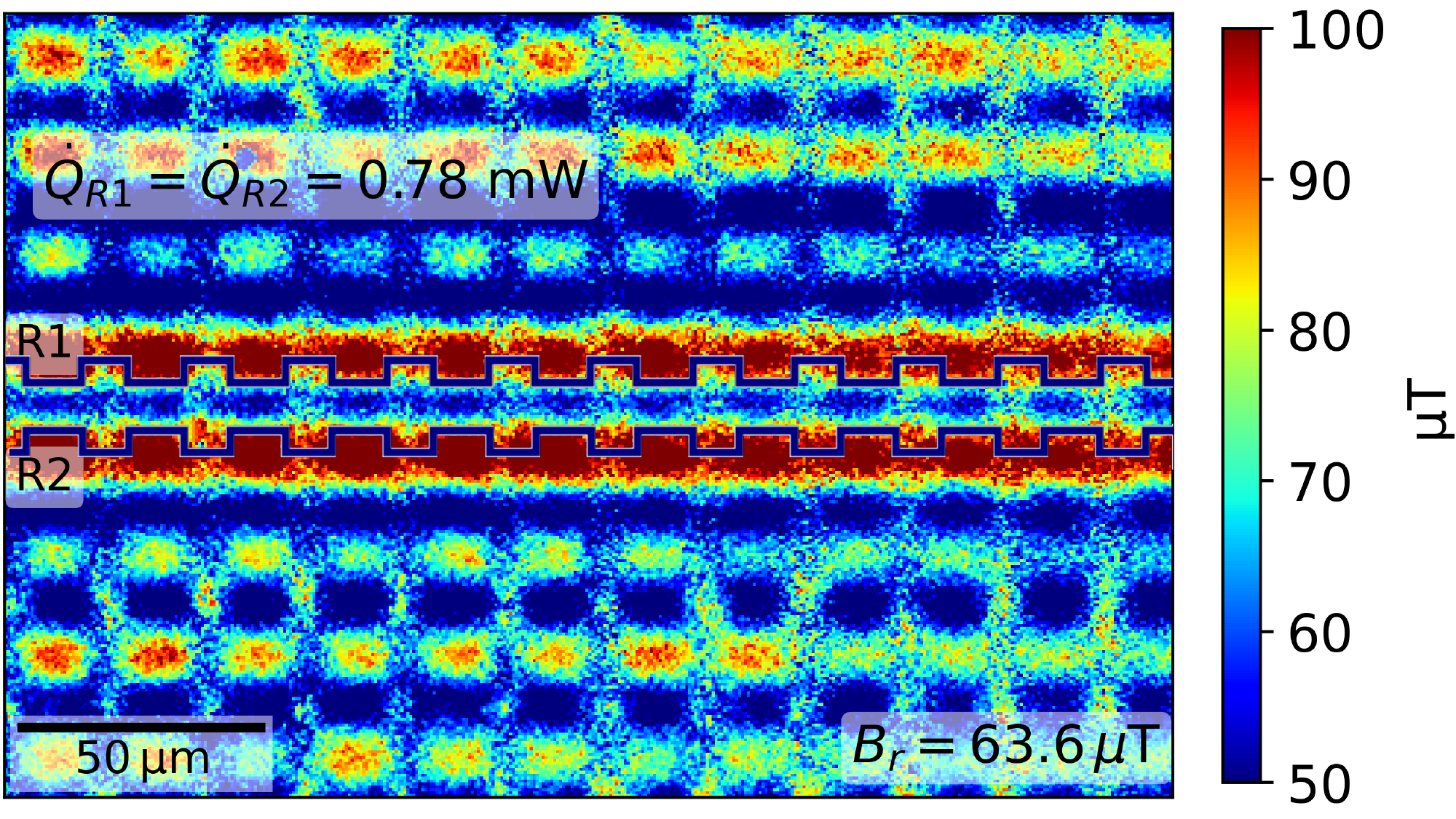}
    \put(-6,52){\footnotesize\textbf{(d)}}
  \end{overpic}

  \caption{(a), (b) Magnetic images for $B_r = 50.8~\upmu\mathrm{T}$ with (a) $\dot{Q}_{R1}=\dot{Q}_{R2}=0$ and (b) $\dot{Q}_{R1}=\dot{Q}_{R2}=0.35$ mW. (c), (d) Magnetic images for $B_r = 63.6~\upmu\mathrm{T}$ with (c) $\dot{Q}_{R1}=\dot{Q}_{R2}=0$ and (d) $\dot{Q}_{R1}=\dot{Q}_{R2}=0.78$ mW. The outer moat resistors used in the experiments are overlaid blue in all four images. In (a) and (c), corresponding to $\dot{Q}_{R1}=\dot{Q}_{R2}=0$, we observe significant amounts of trapped flux near the outer moat resistors; these vortices are circled in white. In (b) and (d), after cooling through $T_c$ with $\dot{Q}_{R1}=\dot{Q}_{R2}\neq0$ and shutting the resistors off at base temperature, the film near the outer moat resistors appears to be flux free and the moats overlaid with the outer moat resistors appearing to trap additional flux. During cooldown, the outer moat resistors act as local temperature maxima, which attract the vortices towards the moats aligned with the resistors, preventing the vortices from trapping in the film. All images and color bars are in units of $\upmu$T. }
  \label{fig:thermal_gradient_flux_mitigation}
\end{figure*}
We evaluated the effectiveness of thermal gradients at mitigating magnetic flux by (1) cooling the chip through $T_c$ with current flowing through the resistors, (2) turning the current off well below $T_c$, and then (3) magnetically imaging the resultant flux distribution. Due to wiring limitations in the microscope, only the outer moat resistors could be evaluated using magnetic imaging. Both designs were subsequently evaluated electrically in the LHe dunk probe, where we observed no significant differences between the effectiveness of each design.

We initially evaluated the effect of heating from a single resistor on flux trapping by cooling the chip with a 10 Hz square wave alternating current flowing through a single outer moat resistor. An alternating current was chosen because we observed a strong dependence of the flux distribution on the direction of the applied direct current, $\Vec{I}_{res}$, relative to the orientation of $\Vec B_r$ (perpendicular to the chip surface), i.e. on the direction of the $\Vec{I}_{res}\times \Vec{B_r}$. This dependence was effectively suppressed using an alternating current during the cooldown (see Appendix~\ref{subsec:DC_expts_discussion}). 

Figure~\ref{fig:single_outer_moat_resistor_measurement}a and b shows magnetic images of the flux distribution at $B_r = 31.6~\upmu\mathrm{T}$ as a function of heater power $\dot{Q}=I_{\mathrm{res}}^2 R$  applied to the outer moat resistors during cooldown, where $I_{\mathrm{res}}$ is the square wave current amplitude and $R$ the heater resistance. The resultant change in the flux in the moats, $\Delta\Phi=\Phi_{\mathrm{on}}-\Phi_{\mathrm{off}}$, is plotted as a function of row index in Fig.~\ref{fig:single_outer_moat_resistor_measurement}c for various values of $\dot{Q}$. $\Delta\Phi$ is calculated as the difference in flux quanta per moat between the current ``on" state, $I_{\mathrm{res}} \neq 0$,  and current ``off" state, $I_{\mathrm{res}} = 0$. In the magnetic images, the field produced by quantized moat flux is also quantized, increasing by $\approx8$ $\upmu$T for every additional $\Phi_0$, allowing the number of flux quanta in each moat to be easily tracked.

As shown in Fig.~\ref{fig:single_outer_moat_resistor_measurement}a, at $\dot{Q} = 0$ and $B_r = 31.6~\upmu\mathrm{T}$ the flux distribution is fairly uniform with most flux trapped in the moats and few vortices trapped in the film near vias $\approx$ 50 $\upmu$m away from the outer moat resistor. As $\dot{Q}$ increases, the moats with outer resistor attract additional vortices, with $\Delta\Phi > 2\Phi_0$ for $\dot{Q} \geq 0.35~\mathrm{mW}$. At $\dot{Q} = 0$, this flux occupied the nearest- and second-nearest-neighbor moat rows. At $\dot{Q} \geq 0.35~\mathrm{mW}$, the heater-induced thermal gradients are sufficient to draw vortices between the moats and flux in the moats toward the resistor moats, as evidenced by the corresponding decrease in flux per moat in the adjacent rows and seen in Fig.~\ref{fig:single_outer_moat_resistor_measurement}a,b. 
\subsection{Flux Trapping Mitigation Using Thermal Gradients}
After characterizing the effect of a single outer moat resistor on the flux distribution, we evaluated the extent to which these resistive heaters could mitigate trapped vortices in the film. For these experiments, we applied equal currents to both outer moat resistors ($I_{\mathrm{res}}=I_{\mathrm{res,R1}}=I_{\mathrm{res,R2}}$, $\dot{Q}=\dot{Q}_{R1}=\dot{Q}_{R2}$) while cooling the sample through $T_c$ under different background fields $B_r$ and values of $I_{\mathrm{res}}$.  In all cases, 10 Hz square-wave alternating currents were applied to each resistor using independent, unsynchronized power supplies. After reaching the setup base temperature, the currents were turned off, and the resulting flux distribution was imaged following thermalization of the chip at $T\approx7$ K.

Fig.~\ref{fig:thermal_gradient_flux_mitigation}a,c show magnetic images at $B_r = 50.8$ and $63.6~\upmu$T, respectively, with the current off during cooldown ($\dot{Q} = 0$). At these relatively high fields which are comparable to Earth's ambient field, vortices populate the film between the moats, including regions near the outer moat resistors (circled in white).

In contrast, Fig.~\ref{fig:thermal_gradient_flux_mitigation}b,d show magnetic images of the same circuit area cooled in the same fields but with the resistor currents on, corresponding to $\dot{Q}=0.35$ and 0.78 mW, respectively. In both cases, flux now concentrates within the moats aligned with the outer moat resistors, while the surrounding film is flux-free. Flux that was previously trapped in the film near the outer moat resistors for $\dot{Q} = 0$, or located in adjacent moats, is now preferentially trapped in the resistor moats.

\begin{figure}[htbp]
    \centering
    \vspace{0.5em}
    \begin{overpic}[width=0.95\linewidth]{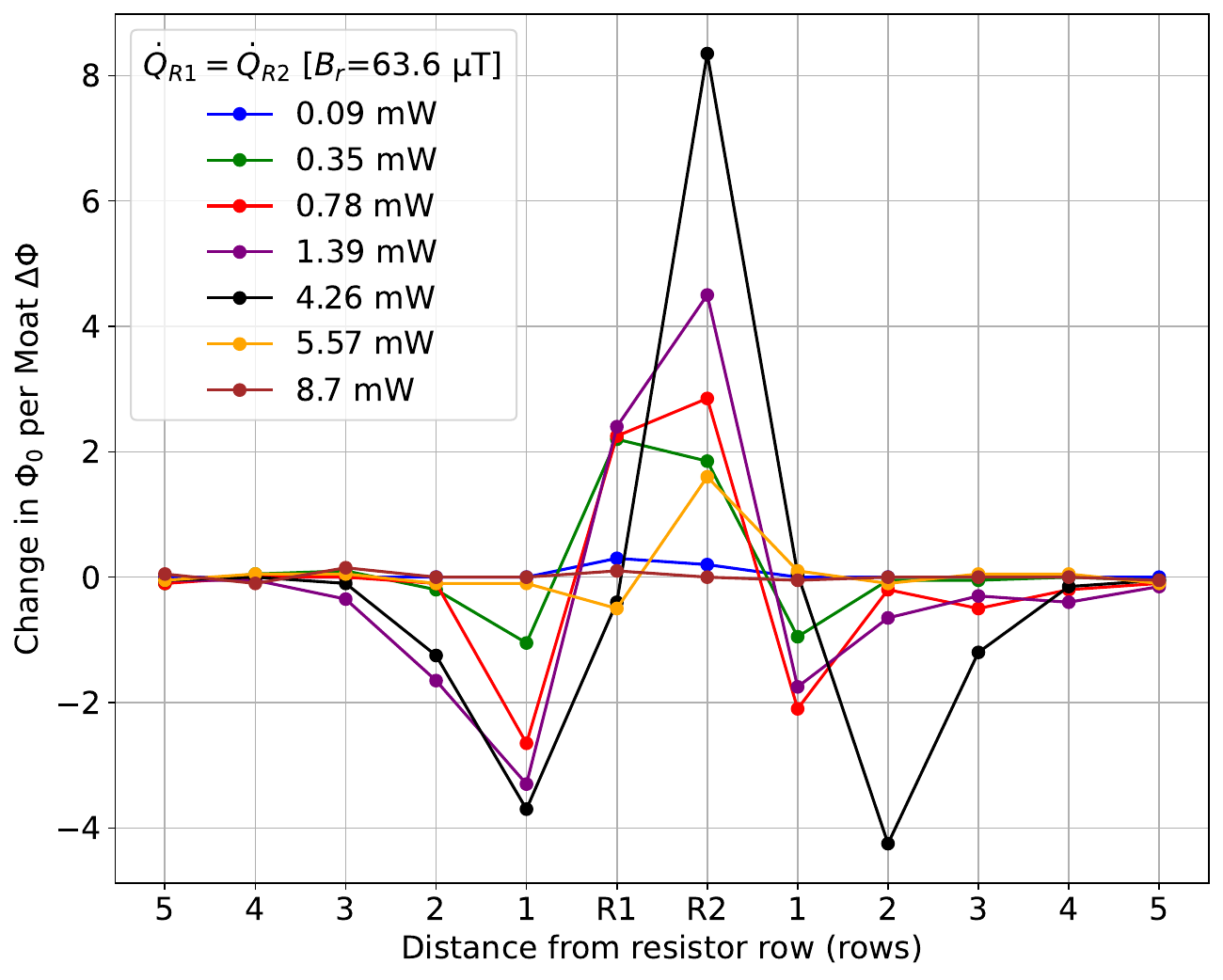}
    \put(-5,68){\footnotesize\textbf{(a)}}
    \end{overpic}
    \vspace{0.5em}
    \begin{overpic}[width=0.95\linewidth]{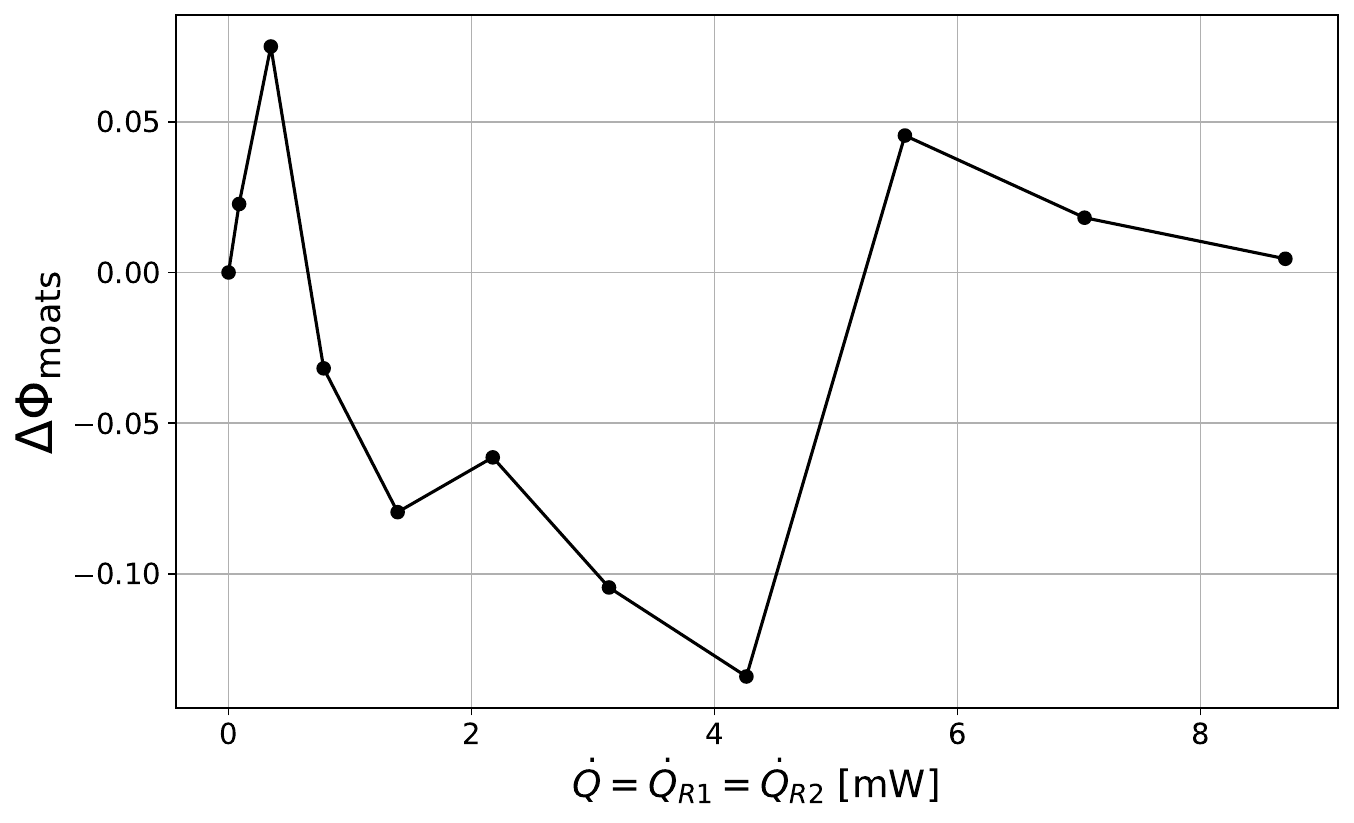}
    \put(-5,68){\footnotesize\textbf{(b)}}
    \end{overpic}

    \caption{(a) Change in the number of flux quanta $\Phi_0$ per moat relative to the $\dot{Q}=\dot{Q}_{R1}=\dot{Q}_{R2}=0$ case, $\Delta\Phi$, for each moat row adjacent to the two outer moat resistor rows R1 and R2 as a function of $\dot{Q}=Q_{R1}=Q_{R2}$ during cooldown at $B_r$ = 63.6 $\upmu$T. (b) Change in total flux in all moats, $\Delta\Phi_{\mathrm{moats}}$, in the microscope field of view (normalized per moat, in units of $\Phi_0$) relative to the $\dot{Q}=\dot{Q}_{R1}=\dot{Q}_{R2}=0$ case,  at $B_r$ = 63.6 $\upmu$T. See text for the discussion of these changes.} 

    \label{fig:average_delta_phi0_vs_qr}
\end{figure}

We quantify the effectiveness of localized heating in mitigating trapped flux by tracking the change in magnetic flux per moat, \(\Delta\Phi\), as a function of applied heater power \(\dot{Q}\). Figure~\ref{fig:average_delta_phi0_vs_qr}a shows \(\Delta\Phi\) for the resistor moats (R1 and R2) and the five neighboring moat rows on both sides of the heaters at \(B_r = 63.6~\upmu\mathrm{T}\).

At low power (\(\dot{Q} = 0.09~\mathrm{mW}\)), we observe a small increase in flux in the R1 and R2 moats, while \(\Delta\Phi \approx 0\) in other moat rows, likely corresponding to the R1 and R2 moats trapping vortices that previously trapped in the nearby film for $\dot{Q}=0$. For \(0.35 \leq \dot{Q} \leq 1.39~\mathrm{mW}\), the trapped flux in the R1 and R2 moat rows increases substantially, with \(\Delta\Phi \geq 2\Phi_0\). In this regime, nearby moat rows exhibit \(\Delta\Phi \leq 0\), indicating that the resistor moats attract flux previously trapped in the adjacent row of moats in the film.

For \(1.39 \leq \dot{Q} \leq 4.26~\mathrm{mW}\), the flux redistribution becomes strongly asymmetric, with vortices preferentially accumulating in the R2 moats and flux redistribution extending to moats as far as four rows away (\(\approx60\)-\(80~\upmu\mathrm{m}\) away). At \(\dot{Q} = 4.26~\mathrm{mW}\), the R2 moats reach \(\Delta\Phi \approx 8\Phi_0\), while the flux change in the R1 moats remain near \(\Delta\Phi \lesssim 0\). This asymmetry likely reflects changes in the on-chip temperature profile, which may be influenced by asymmetries in the surrounding circuitry. In particular, the R1 row is located closer to neighboring passive resistors, which likely act as thermal sinks, reducing local temperature and shifting the peak temperature toward the R2 row, resulting in preferential flux trapping in the R2 moats; see Fig.~\ref{fig:passive_gds}b.

For \(\dot{Q}\geq 5.57~\mathrm{mW}\), local heating becomes significant, and the Nb film in the resistor region and nearby areas likely become non-superconducting when the resistors are on and differences relative to the \(\dot{Q}=0\) case likely arise from transient dynamics during the brief time interval after the resistors are turned off and the film becomes fully superconducting. At \(\dot{Q} \geq 8.7~\mathrm{mW}\), changes in the moat flux become negligible, consistent with the resistors heating the entire film above \(T_c\).

Changes in the total flux in all moats in the area of observation offers an additional insight into the flux redistribution dynamics induced by the localized heating. Initially, the overall flux in the moats (normalized per moat) increases, $\Delta\Phi_{\mathrm{moats}} > 0$, as vortices that were trapped in the film near the resistors at $\dot{Q}=0$ trap in the heated moats at $\dot{Q} \neq 0$. As $\dot{Q}$ increases, the overall flux reduces, $\Delta\Phi_{\mathrm{moats}} < 0$. We believe this reduction is caused by the spatial extent of the thermal gradient increasing and enabling flux expulsion to the edges of the film or outside the field of view. This trend continues until local heating above $T_c$ starts to occur at $\dot{Q}=4.26$ mW, where the behavior reverts back to $\Delta\Phi_{\mathrm{moats}} > 0$ and likely depends on transient thermal gradients that form when the resistors are turned off. At $\dot{Q}>8.7$ mW, $\Delta\Phi_{\mathrm{moats}} \approx 0$, as the resistors heat the entire film above $T_c$ (see Fig.~\ref{fig:average_delta_phi0_vs_qr}b).
\subsection{Electrical Measurements of the Circuit}
The presented magnetic images indicate that local heating can mitigate trapped flux. While the images likely probe only vortices in the M7 ground plane, vortex behavior is expected to be similar in M4 because the two ground planes are symmetric about the resistors. Meanwhile, flux trapping in M1 has previously been found to have a negligible impact on circuit performance~\cite{shregNew}. Moreover, the moat measurements are sensitive to the flux distribution across all three ground planes. For example, at $B_r=63.6$ $\upmu$T and $\dot{Q}=0.09$ mW, the moats aligned with R1 and R2 exhibit $\Delta\Phi=0.25\Phi_0$, while all other moats have $\Delta\Phi=0$ (Fig.~\ref{fig:average_delta_phi0_vs_qr}), indicating that an additional $0.25\Phi_0$ per moat that was trapped in the superconducting film for the $\dot{Q}=0$ case is instead captured by the moats when heating is applied. In contrast, the images show only $\approx0.1\Phi_0$ per resistor moat removed from the film, implying that additional flux redistribution occurs in the lower ground planes that are not directly visible (Appendix~\ref{subsec:supp_images} and Fig.~\ref{fig:100uA_63_6uT}). Nevertheless, electrical characterization of functional circuitry provides a more direct assessment of whether controlled local thermal gradients improve circuit performance.

We verified the impact of localized heating on electrical performance by measuring the critical current, $I_c$, of the JJ chain between the outer moat resistors in the NV-diamond microscope, as well as in a separate electrical test setup using a LHe immersion test probe. Josephson junction chains were selected as model circuitry because the critical current of individual junctions in the chain can serve as magnetic field and temperature sensors, and be measured using just a few wires.

The critical current $I_c$ in JJ chains was found to be relatively insensitive to the cooldown magnetic field $B_r$ and to flux residing outside the junction layer (Performance of more complex circuits such as shift registers can fail in fields as low as $B_r\leq$ 1 $\upmu$T~\cite{shregNew,shregOld}). For example, in the images presented in Fig.~\ref{fig:thermal_gradient_flux_mitigation}a,c, the JJ chain $I_{c}$ is very similar to the $B_r$ = 0 case despite a significant amount of vortices located in the ground planes (at various distances above and below the JJs in image). That is, perpendicular vortices located in the ground planes of the circuit in proximity to the JJs do not affect the critical current of the JJs, in agreement with previous observations~\cite{Kirtley2019,tinkham2004introduction,PhysRevLett.81.204}. To induce a measurable degradation in $I_c$, we cooled the device with an applied current $I_{s} \gg I_c$ to the JJ chain and then turned the current off well below $T_c$. This procedure generates a local magnetic field leading to flux trapping near and inside the junctions, and to a corresponding suppression of $I_c$, thereby providing a degraded operating condition against which the effectiveness of the localized heating can be evaluated.

\begin{figure}[htbp]
    \centering

    \begin{overpic}[width=0.95\linewidth]{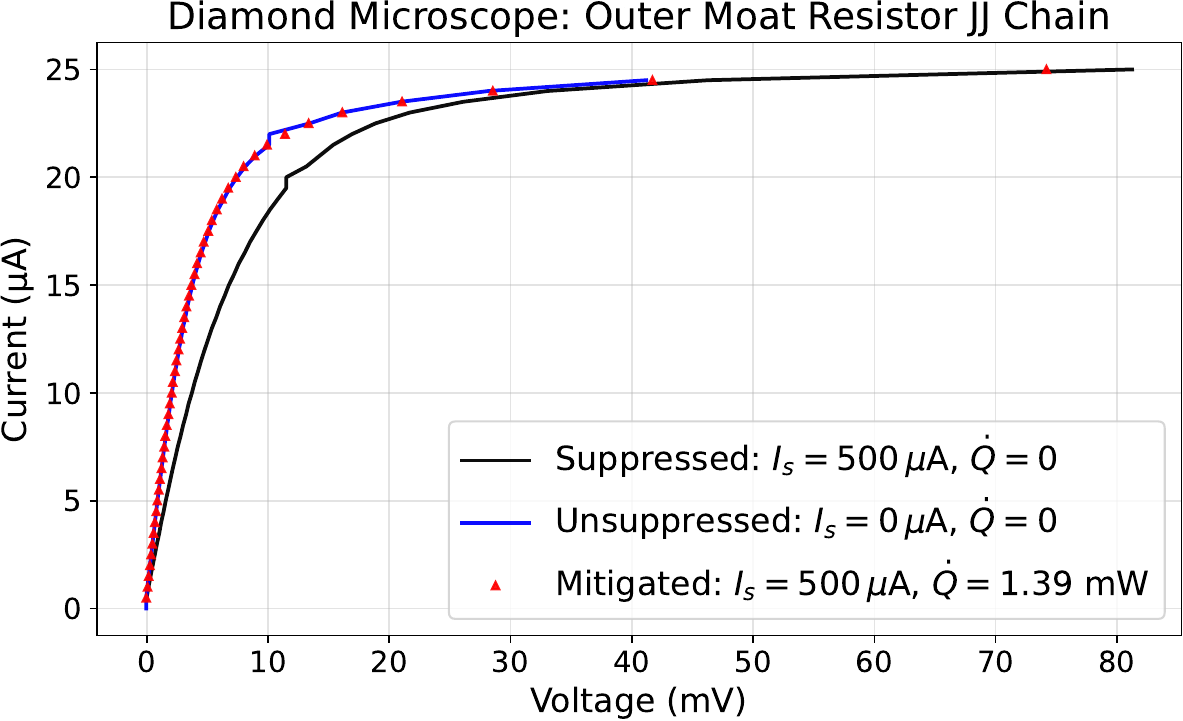}
        \put(-5,58){\footnotesize\textbf{(a)}}
    \end{overpic}

    \vspace{1.2em}

    \begin{overpic}[width=0.95\linewidth]{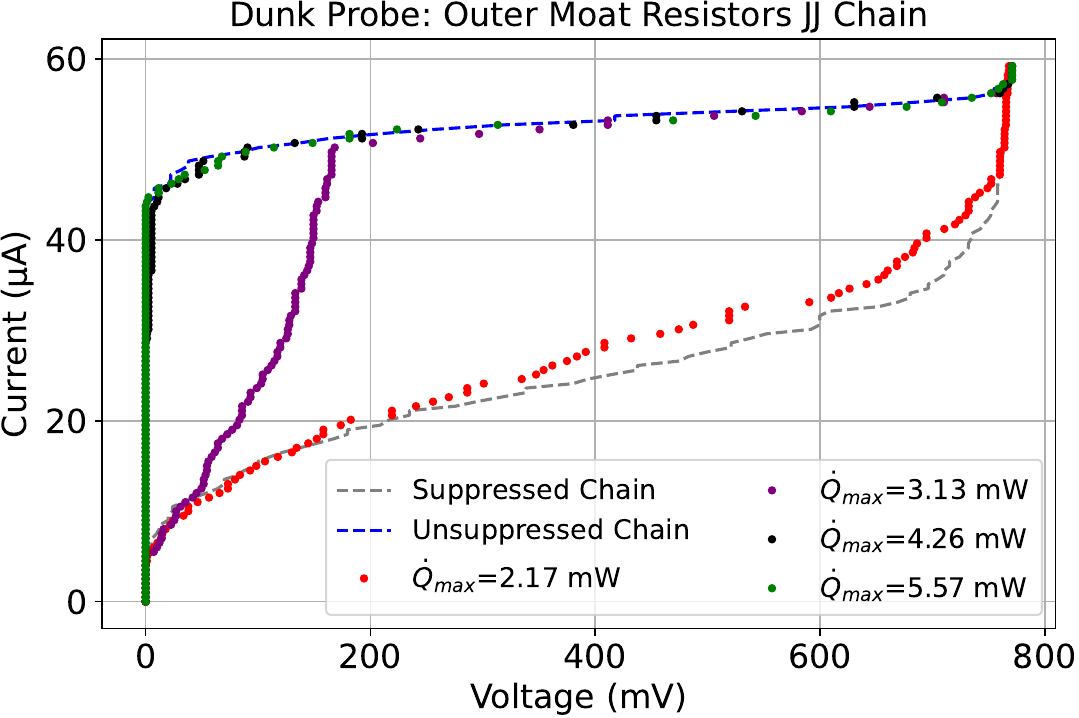}
        \put(-5,58){\footnotesize\textbf{(b)}}
    \end{overpic}

    \vspace{1.2em}

    \begin{overpic}[width=0.95\linewidth]{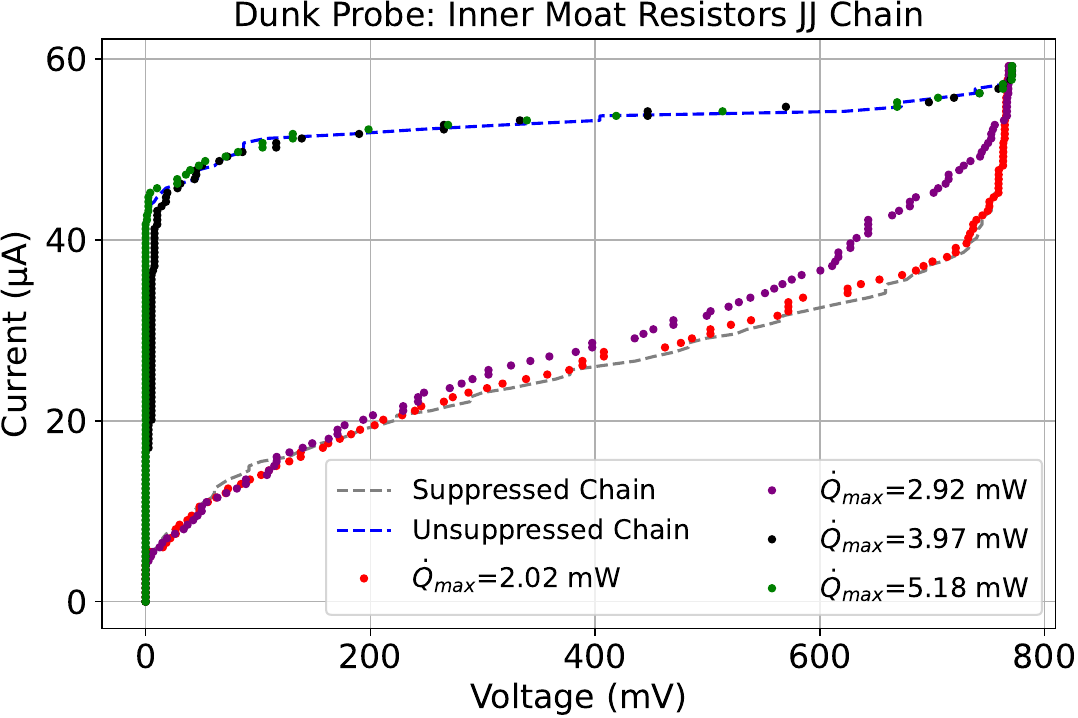}
        \put(-5,58){\footnotesize\textbf{(c)}}
    \end{overpic}

    \caption{(a) Two-point IV curves of the JJ chains, 300 Josephson junctions in series, taken in the NV diamond microscope apparatus; series resistance of the leads contribute to the initial slope of the IV cureve. Cooling through $T_c$ with $I_s=500$ $\upmu$A suppresses the IV behavior. This behavior is mitigated by cooling through $T_c$ with $\dot{Q}=1.39$ mW applied to both outer moat resistors surrounding the chain (see Appendix~\ref{subsec:supp_images} for images of this behavior). (b)-(c) Four-point IV curves taken of the JJ chains surrounding the outer moat resistors (b) and inner moat resistors (c) in the LHe dunk probe. The chains were suppressed by cooling through $T_c$ with $I_s=1.5$ mA, and this suppression was mitigated by pulsing both the outer (b) or inner (c) moat resistors to $\dot{Q}_{max}$. The chains recover their unsuppressed behavior for $\dot{Q}_{max}\gtrsim$ 3.97 mW. }

    \label{fig:IC_measurements}
\end{figure}

We initially performed $I_c$ measurements in the NV-diamond microscope, where the critical current of the JJ chain between the outer moat resistors was intentionally suppressed and then evaluated as a function of $\dot{Q}$ during cooldown, consistent with the conditions used for magnetic imaging. To induce degradation of $I_c$, the chip was cooled with current $I_s=500\,\mathrm{\upmu A} \gg I_c$ applied to the JJ chain. This produces a measurable suppression of the IV curve, as shown in Fig.~\ref{fig:IC_measurements}a. We then repeated the cooldown with both $I_s=500\,\mathrm{\upmu A} \gg I_c$ applied to the JJ chain and a current $I_{\mathrm{res}}$ applied to the nearby outer moat resistors, corresponding to $\dot{Q}=\dot{Q}_{R1}=\dot{Q}_{R2}=1.39$ mW. In this case, the IV curve recovers its baseline performance, indicating that the suppression induced by $I_s$ is mitigated by the resistive heating during cooldown (see Appendix~\ref{subsec:supp_images} for images of this effect). The NV-diamond microscope measurements, however, were limited by noise and wire count, as the system was initially designed for optical-only measurements, with in-situ electrical upgrades still ongoing (see Appendix~\ref{subsec:microscope_active_upgrade}).

We further validated the effectiveness of thermal gradients in mitigating trapped flux by testing the chip LHe dunk probe. While the dunk probe did not have a bias coil to vary the background magnetic field, we performed all cooldowns and measurements without any magnetic shields in the setup,  corresponding to $B_r \approx$ 50 $\upmu$T from the Earth's magnetic field. 

In the dunk probe, rather than focusing solely on mitigating flux during cooldown, we considered a more general case in which the film is already superconducting and magnetic flux is trapped in the circuit. Such conditions can arise if the thermal gradients applied during cooldown are insufficient to remove all trapped flux, or if magnetic fields generated during circuit operation produce additional vortices. Instead of thermally cycling the entire chip, we investigated whether pulsing heat locally can mitigate trapped flux, enabling continuous circuit operation while removing parasitic flux in the vicinity of the circuitry.

In the experiments presented in Fig.~\ref{fig:IC_measurements}b and c, the chip was cooled with a current $I_{\mathrm{s}} = 1.2~\mathrm{mA}$ applied to the JJ chain and DC $I_{\mathrm{res}} = 1.1~\mathrm{mA}$ applied to the resistors to maximize the amount of vortices that form in the film. In the suppressed case, the critical current is reduced to $<5~\upmu\mathrm{A}$ for all heater designs, compared to its nominal value of $\approx 40~\upmu\mathrm{A}$. The magnetic field created by  $I_{\mathrm{s}}$ ($B_{\mathrm{s}} \approx 240/r\,\upmu\mathrm{T}$, where $r$ is the distance from the chain in $\upmu\mathrm{m}$) during cooldown likely generates vortices throughout the chain, leading to $I_{\mathrm{c}}$ suppression in the majority of the junctions. To reverse this effect without warming and re-cooling the entire chip, the nearby resistors were turned on for 1 second and then slowly ramped down 10 $\upmu$A every 50 ms.

As seen in Fig.~\ref{fig:IC_measurements}b and c, the number of JJs with suppressed $I_c$ changes as a function of $\dot{Q}_{max}$. We can estimate this number using the voltage at $I=40$ $\upmu$A, just below the nominal critical current of the junctions. After the cooling, more that 95$\%$ of the junctions have suppressed $I_c$. After applying current to the outer moat resistors, corresponding to $\dot{Q}_{max}=2.17$, 3.13, 4.26, and 5.57 mW, $\approx$ 6\%, 80\%, 99\%, 100\% of JJs in the chain recovered baseline critical current, respectively. For the inner moat resistors, $\dot{Q}_{max}=2.02$, 2.92, 3.97, and 5.18 mW resulted in $\approx$ 1\%, 15\%, 99\%, 100\% of JJs in the chain recovering the baseline IV characteristics, respectively.

\subsection{Local Temperature Characterization}
We characterize the local temperature at the JJ chains using the temperature dependence of the Josephson junction gap voltage, \(V_{\mathrm{gap}} = 2\Delta/e\), where \(\Delta\) is the superconducting energy gap and \(e\) is the electron charge. Using the approximate BCS relation for the temperature dependence of the superconducting gap and inverting for temperature gives
\begin{equation}
T = T_c \left[
\left(
\operatorname{arctanh}\!\left(
V_{\mathrm{gap}} / V_{\mathrm{gap},0}
\right) / 1.74
\right)^2 + 1
\right]^{-1}
\label{eq:vgap_vs_temp}
\end{equation}
\cite{tinkham2004introduction,RevModPhys.76.411}. The zero-power gap voltage, $V_{\mathrm{gap},0} = 2.67~\mathrm{mV}$, was measured using the He bath temperature $T_0 = 4.2~\mathrm{K}$ and critical temperature $T_c = 9.25~\mathrm{K}$ and previous measurements of similar circuits~\cite{9769893}.

\begin{figure}[htbp]
    \centering
    \vspace{0.5em}
    \begin{overpic}[width=0.95\linewidth]{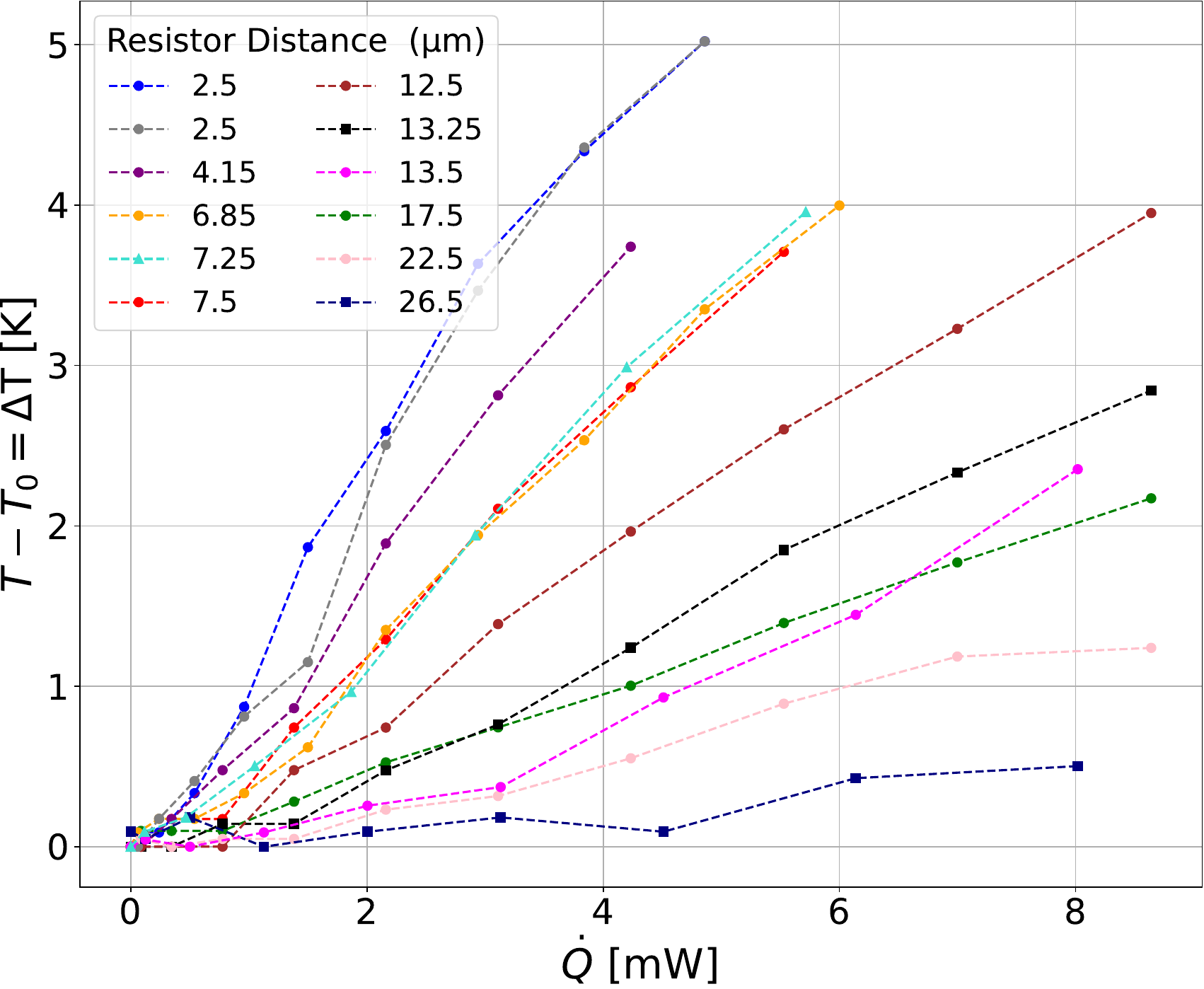}
    \end{overpic}
    \caption{Measured $\Delta T$ vs $\dot{Q}$ in the LHe dunk probe as a function of distance from resistor. Experiments were performed by measuring the gap voltage of the JJ chains as current was applied to a single resistor (see Eq~\ref{eq:vgap_vs_temp}). Measurements were performed with the straight (circles), outer moat (squares), and inner moat (triangles) resistors and JJ chains 2-4 and 6 (see Fig.~\ref{fig:passive_gds}b) }
    \label{fig:temp_characterization}
\end{figure}

In Fig.~\ref{fig:temp_characterization}, the extracted temperature as a function of resistor power is plotted for various distances between the junction chain and the heater. In these measurements taken in the LHe dunk probe, we applied current to a single straight, inner, or outer moat resistor and measured the gap voltage at nearby JJ chains, specifically JJ chains 2-4 and 6~(see Fig.~\ref{fig:passive_gds}b). For inner and outer moat resistors, the distance was taken as the average distance of the resistor to the JJ chain. 
\section{Discussion}
\subsection{Magnetic Images}
The magnetic imaging indicates that local thermal gradients can prevent magnetic flux from trapping in nearby circuit areas in field-cooled experiments. Fig.~\ref{fig:thermal_gradient_flux_mitigation}b and d show that cooling the circuit in ambient magnetic fields with current applied to the resistors prevents vortex formation in the film between the adjacent resistors, as well as in the film regions on either side of the resistors. 

During cooldown through $T_c$ with current applied to the resistors, the circuit initially reaches a state in which Nb ground plane films above and below the resistors are in the normal state, while the adjacent circuitry regions are superconducting but very close to $T_c$. As cooling continues, the entire chip gradually transitions into the superconducting state with the circuitry regions being slightly colder than in the resistors region. In both regimes, vortices nucleated in the circuitry regions are attracted to the warmer regions, where their energy is lower, as any competing pinning forces $F_{\mathrm{pin}}$ are weak near $T_c$, scaling approximately as $(1 - T/T_c)$~\cite{dew1974flux,kapur2026mitigationmagneticfluxtrapping,tinkham2004introduction}. As the film continues to cool below $T_c$, the competing $F_{\mathrm{pin}}$  increases and becomes comparable to and eventually exceeds the thermal force, $F_{\mathrm{th}}$, and the moat attraction force. At these lower temperatures, vortex nucleation and motion is suppressed as the vortex energy and $F_{\mathrm{pin}}$ increase significantly as $T_c-T$ increases. Consequently, vortices that are driven toward the resistor regions near $T_c$ become immobilized and remain trapped there upon further cooling.

The net result is that vortices which would otherwise pin in the film or occupy adjacent moats for $\dot{Q}=0$ are instead driven toward the hot resistor moats for $\dot{Q} \neq 0$, where they remain trapped even after the heaters are turned off. In the most extreme cases, we observed an additional 8$\Phi_0$ trapping in moats, corresponding to a total flux of $16\Phi_0$ trapped in a single moat.
\subsection{Electrical Measurements}
The extent at which local thermal gradients can suppress trapped flux in the film can be quantified using the temperature measurements performed in the LHe dunk probe. The results likely do not apply directly to the NV diamond microscope measurements, as the  cooling conditions and chip packaging differ significantly. However, superconducting circuits in practical deployment environments are more closely represented by the LHe dunk probe configuration than by the NV microscope setup. 

The resistors are modeled as narrow-line linear heaters running in the y-axis direction, $\dot Q\delta(x-x_i)$, where $x_i$ are locations of the resistors, as the distance to the Josephson junctions serving as temperature sensors ($\lesssim 100$ $\upmu$m) and the width of the resistors (1 $\upmu$m) are much smaller than the resistor length (3 mm). Under this approximation, the heat flow is treated as one-dimensional in the $x$-axis direction, perpendicular to the heaters. Heat is continuously removed in the z-direction perpendicular to the chip surface through the multilayer chip and into the liquid helium bath.

Away from the resistor ($|x-x_i|>0.5$ $\upmu$m), the steady-state heat diffusion equation describing the temperature profile is
\begin{equation}
    \frac{d}{dx}\left(k t \frac{dT}{dx}\right)-G(T-T_0)= 0,
\end{equation}
where $T(x)$ is the local temperature, $T_0$ is the bath temperature, $k$ is the effective chip thermal conductivity ($\mathrm{W\,m^{-1}\,K^{-1}}$), $t$ is the effective thickness of the chip region participating in thermal transport ($\mathrm{m}$), and $G$ is the effective vertical thermal conductance per unit length to the LHe bath ($\mathrm{W\,m^{-1}\,K^{-1}}$). The first term represents lateral heat conduction, while the second term represents local heat loss from the film stack into the substrate and helium bath. The boundary conditions to this heat equations are ${T(|x| \rightarrow \infty) = T_0}$, the bath temperature. The local temperature rise on the resistors $\Delta T(x_i)$ is proportional to the applied heat power $\Delta T(x_i)=\dot Q R_{\mathrm{bath}}$, where $R_{\mathrm{bath}}$ is the effective thermal resistance between the heater and LHe bath.

Assuming $k$, $t$, and $G$ are approximately constant over the measured temperature range  and defining $\Delta T(x)=T(x)-T_0$ gives, 
\begin{equation}
    \frac{d^2\Delta T}{dx^2}-\frac{\Delta T}{\lambda_{\mathrm{th}}^2}=0,
\end{equation}
where $\lambda_{\mathrm{th}}=\sqrt{kt/G}$ is the thermal healing length.  The solution to this equation with the described boundary conditions is:
\begin{equation}
    \Delta T(x) = \dot Q R_{\mathrm{bath}}\Sigma_i e^{-|x-x_i|/\lambda_{\mathrm{th}}},
    \label{eq:basic_exp_model}
\end{equation}
valid at distances greater than the heater width; see also~\cite{10.1063/1.1663912,dane2022self}.
\begin{figure}[htbp]
    \centering
    \vspace{0.5em}
    \begin{overpic}[width=0.95\linewidth]{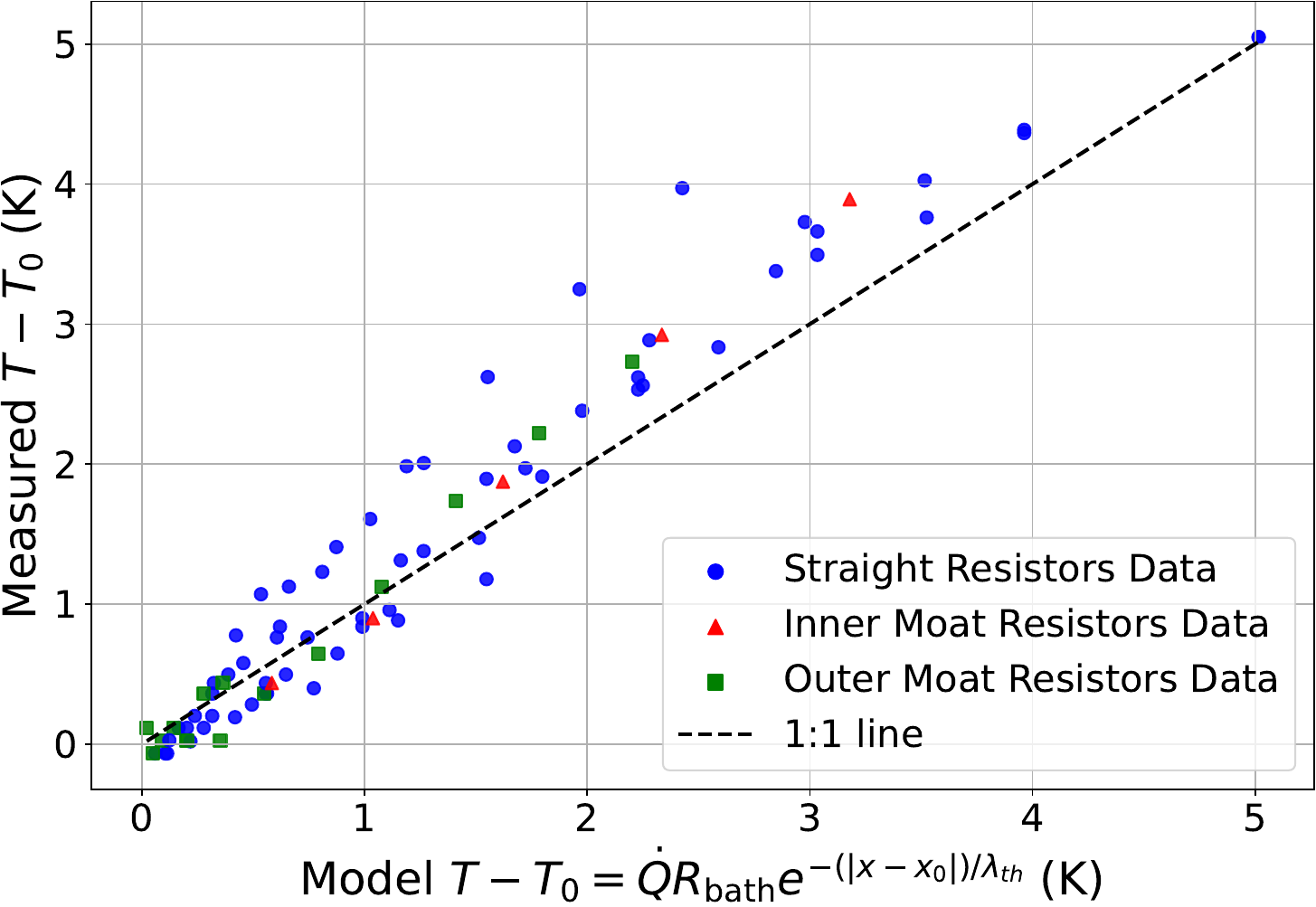}
    \end{overpic}
    \caption{Measured $\Delta T$ vs predicted $\Delta T$ from Eq.~\ref{eq:basic_exp_model}, where $R_{\mathrm{bath}}$ and $\lambda$ are fitted parameters. The black dashed line shows ideal agreement between the model and empirical data. The model shows good agreement with the data and the fitted $R_{\mathrm{bath}}=1430\pm70$ K/W and $\lambda=7.7\pm0.5$ $\upmu$m values are consistent with independent calculations of these values in Appendices~\ref{subsec:thermal_healing_length} and~\ref{subsec:circuit_thermal_RC}.}
    \label{fig:fitted_exp_model}
\end{figure}

We fit the single resistor data in Fig.~\ref{fig:temp_characterization} to Eq.~\ref{eq:basic_exp_model} to extract parameters $\lambda_{th}$ and $R_{\mathrm{bath}}$. The dashed black line in Fig.~\ref{fig:fitted_exp_model} shows the predicted temperature rise at the locations of the Josephson junctions serving as local thermometers in Fig.~\ref{fig:temp_characterization} at $R_{\mathrm{bath}} = 1430\pm70$ K/W  and $\lambda_{\mathrm{th}}=7.7\pm0.5$ $\upmu$m. The model provides good agreement with the measured data with deviations likely resulting from the imperfect assumption that $k$ and $G$ are constant over the temperature range. The estimates of $\lambda_{\mathrm{th}} = \sqrt{k t / G}$ and $R_{\mathrm{bath}}$ using the estimated values of the thermal conductivity and interface resistances of the circuit materials are given  in Appendices~\ref{subsec:thermal_healing_length} and~\ref{subsec:circuit_thermal_RC} and are in general agreement with fitted values.

\begin{figure*}[htbp]
    \centering

    \begin{minipage}{0.48\textwidth}
        \centering
        \begin{overpic}[width=\linewidth]{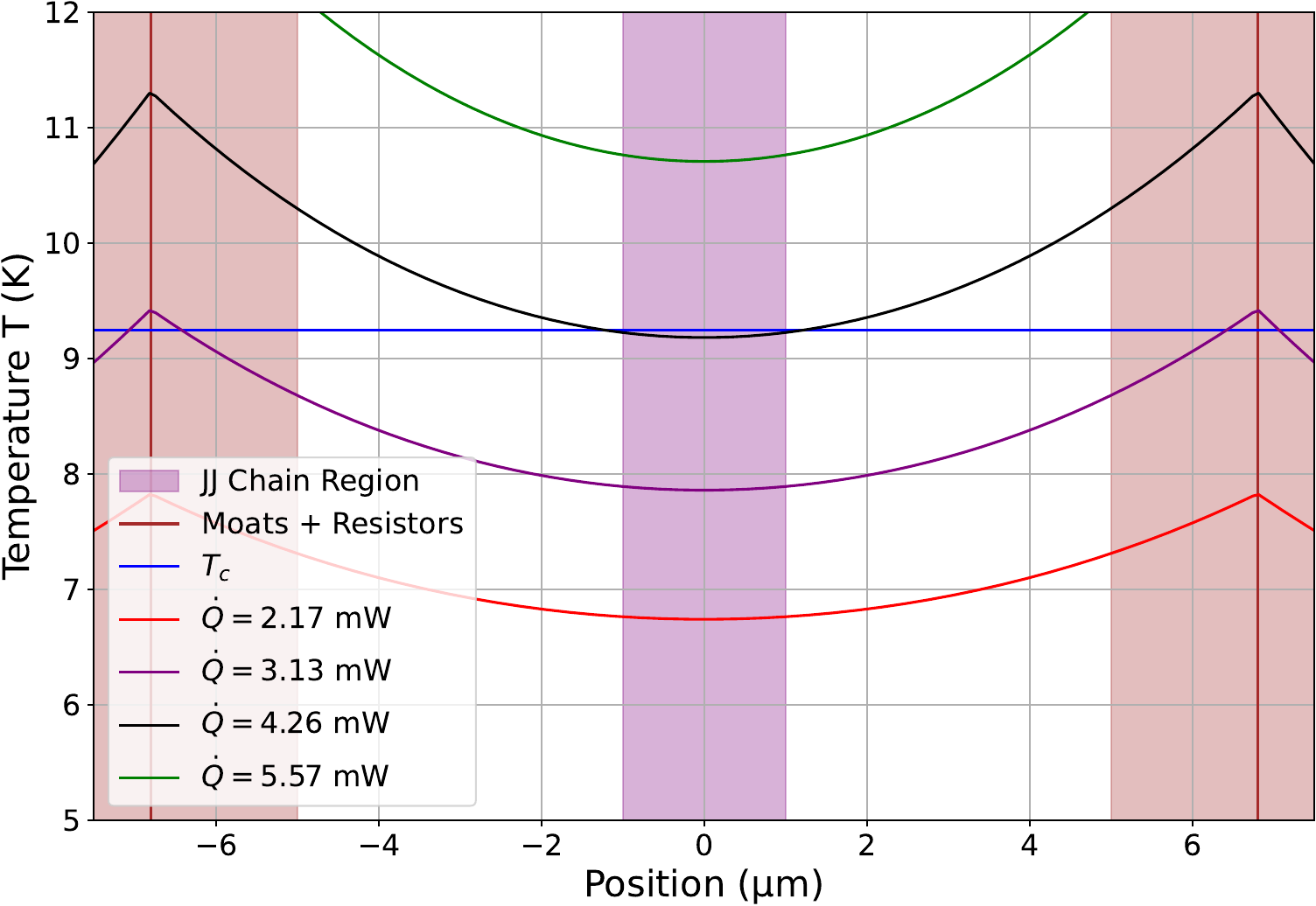}
            \put(-5,68){\footnotesize\textbf{(a)}}
        \end{overpic}
    \end{minipage}
    \hfill
    \begin{minipage}{0.48\textwidth}
        \centering
        \begin{overpic}[width=\linewidth]{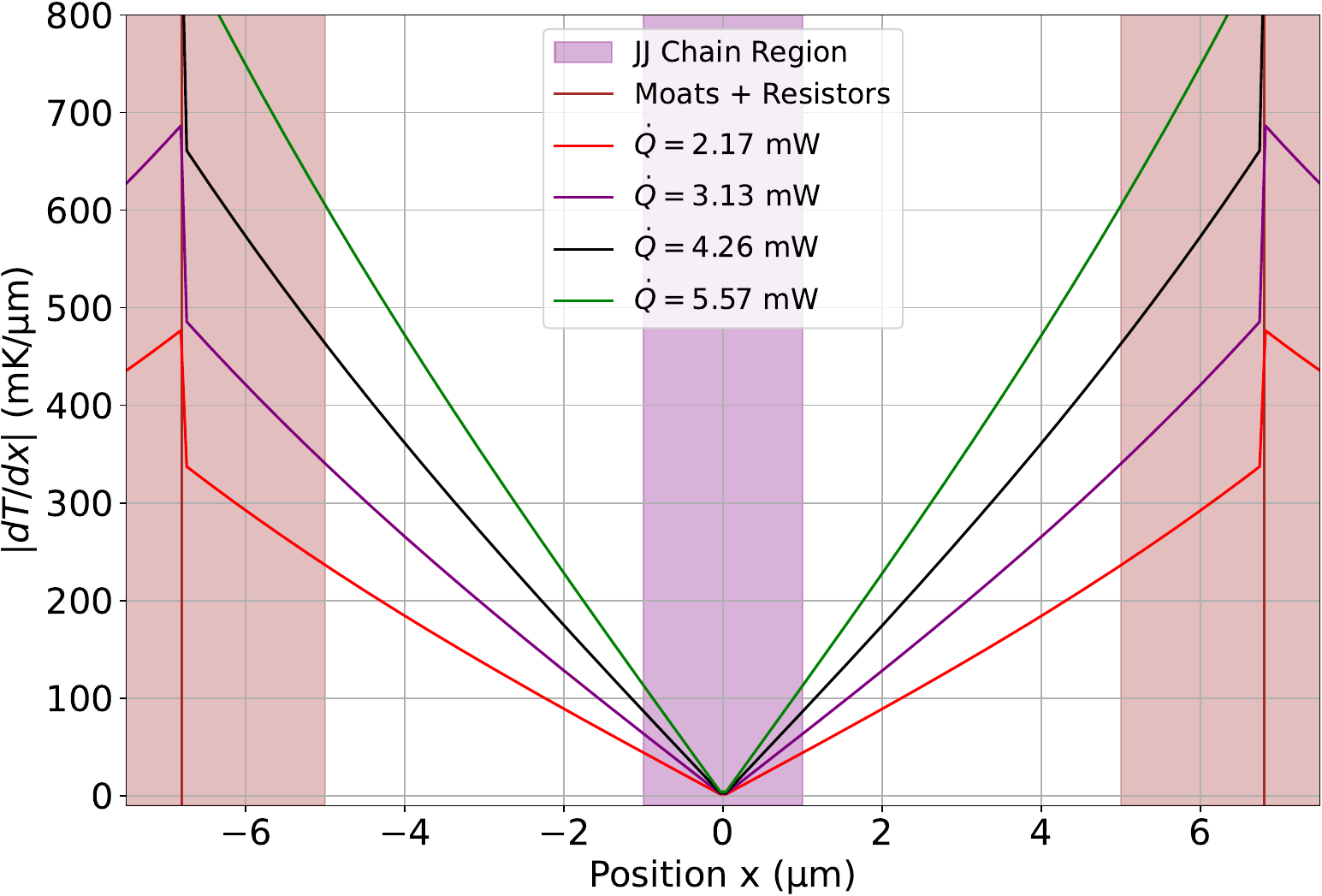}
            \put(-5,68){\footnotesize\textbf{(b)}}
        \end{overpic}
    \end{minipage}

    \caption{Modeled (a) temperature and (b) $|dT/dx|$ profiles between the two outer moat resistors. The JJ chain serving as the local thermometer is centered at $x=0$ and occupies the violet-colored region.  We modeled the profiles at the same values of $\dot Q$ as were used for flux mitigation in Fig.~\ref{fig:IC_measurements}b. At the JJ array location ($x=\pm1$ $\upmu$m) and the moat resistors powers $\dot{Q}_{max}=2.17$, 3.13, 4.26, and 5.57 mW, the corresponding average $T=$ 6.75, 7.87, 9.20, and 10.72 K; and average $|dT/dx|=$ 22, 32, 44, and 57 mK/$\upmu$m. In electrical measurements, these corresponded to 6\%, 80\%, 99\%, and 100\% of JJs in the chain recovering the baseline behavior, respectively. }
    
    \label{fig:thermal_modeling_ab}
\end{figure*}

In Fig.~\ref{fig:thermal_modeling_ab}a and b, the temperature and $\nabla T$ profile of the outer moat resistors based on the derived temperature model is plotted for the four resistor powers used in the electrical measurements presented in Fig.~\ref{fig:IC_measurements}b. We approximate the outer moat resistors as a straight line whose distance from the JJ chain is the average distance of the outer moat resistors to the JJ chain. We evaluate outer moat resistors as this resistor array was tested in the NV diamond microscope and LHe dunk probe, although we do not observe any appreciable differences in flux trapping behavior across resistor designs (see Fig.~\ref{fig:IC_measurements}b,c).

As shown in Fig.~\ref{fig:IC_measurements}b, partial mitigation of the suppressed JJ chain is observed at $\dot{Q}_{\mathrm{max}} = 2.17$ and $3.13~\mathrm{mW}$. At these heating powers, the JJ chains reach temperatures of 6.7 and 7.8~K, with corresponding average thermal gradients in the JJ chain region of 22 and 32~mK/$\upmu$m, respectively. Based on the IV characteristics, these heating powers restore approximately 6\% and 80\% of the junctions in the chain to their unsuppressed state, respectively. 

Based on Fig.~\ref{fig:IC_measurements}b, fully restoring $\geq$~99\% of the JJs to the unsuppressed state requires $\dot{Q}_{\mathrm{max}} \geq 4.97~\mathrm{mW}$. This is consistent with the modeled temperature profiles shown in Fig.~\ref{fig:thermal_modeling_ab}a,b, where $\dot{Q}_{\mathrm{max}} = 4.97~\mathrm{mW}$ corresponds to heating the JJ chain to 9.2 K and produces an average over the chain region  thermal gradient of 44 mK/$\upmu$m. At this heating power, the resistors locally drive the majority of the film outside the JJ chain above $T_c$, while keeping the JJ chain $T_c-T\approx50$ mK. At these temperatures, despite the relatively weak thermal gradient, the thermal force is stronger than any competing forces, namely the pinning force, resulting in the JJ chain becoming flux free, except a few junctions. The unsuppressed behavior is fully recovered when increasing $\dot{Q}_{\mathrm{max}}$ even further, as the entire film is heated above $T_c$. This behavior is analogous to the chip thermal cycling and not due to temperature gradient effects.
\subsection{Thermal Forces During Field Cooling}
As discussed, the JJ chains are relatively insensitive to vortices formed from $B_r$, making it difficult to directly assess mitigation of vortices formed from field cooling in $B_r\neq0$ in the electrical measurement setup. To address this, we use the thermal model to simulate the effects of field cooling, in which the heater power is ramped down slowly after reaching a $\dot{Q}_{max}$ that locally heats the circuitry region above $T_c$, as performed in the electrical measurements for $\dot{Q}_{\mathrm{max}} \geq 5.57~\mathrm{mW}$. In these experiments, the heating power was reduced in steps of 0.9 $\upmu$W every 50~ms. After each step, the system can be approximated as being in steady state, as the thermal RC time constant is $\lesssim$ 1 ms; see Appendix~\ref{subsec:circuit_thermal_RC}.

Vortex energy  in the film strongly depends on the magnetic field penetration depth and, hence, on temperature:

\begin{equation}
\begin{split}
\varepsilon_{\mathrm{v}}
&=
\frac{\Phi_0^2 }
     {4\pi \upmu_0 \lambda(T)^2}
(\ln\!
\frac{\lambda(T)}{\xi(T)}
+C),
\end{split}
\label{eq:vortex_energy}
\end{equation}
where the energy is the per unit length, $C \approx 0.4$ is a constant related to the vortex core energy, \(\Phi_0\) is the magnetic flux quantum, \(\upmu_0\) is the vacuum permeability, \(\lambda(T)\) is the temperature-dependent London penetration depth, and $\xi(T)$ is the temperature-dependent coherence length. A thermal gradient creates a gradient of the vortex energy corresponding to a thermal force, $\vec{F_\mathrm{{th}}}=-\vec{\nabla} \varepsilon_{\mathrm{v}}$, pushing vortices into the warmer regions of the film where their energy is lower than in the colder regions. 
Using the thermal model in Eq.~\ref{eq:basic_exp_model}, we calculated the local temperature and temperature gradient in the circuitry region, defined as $\pm 1~\upmu$m about the midpoint between the two heaters as a function of $\dot{Q}$ (see Fig.~\ref{fig:thermal_modeling_ab}). This region has the lowest temperature and the lowest gradient due to its symmetric position between the heaters. Vortex nucleation is expected in this region, as it is the first part of the heated structure to enter the superconducting state. Any vortex that forms outside this region, closer to the heaters, will experience a larger temperature gradient and a larger $F_{\mathrm{th}}$.

In Nb films, the temperature dependence of the penetration depth follows closely the two-fluid model $\lambda=\lambda_0(1-t^4)^{-1/2}$ while the coherence length follows the Ginzburgh-Landau expression  $\xi(T)=\xi_0(1-t)^{-1/2}$ from about 2 K to $T_c$, where $t=T/T_c$.  From Eq.~\ref{eq:vortex_energy}, the thermally induced force on a vortex per unit length is given by
\begin{equation}
\begin{split}
F_{\mathrm{th}}
&=
\frac{\Phi_0^2 \nabla T}
     {4\pi \upmu_0 \lambda_0^2 T_c}
\Biggl[
4t^3 \ln\!\left(
\frac{ \beta \lambda_0}{\xi_0}
\sqrt{
\frac{1}{(1+t)(1+t^2)}
}
\right)
\\
&\qquad
+
\frac{
(1+t)(1+t^2)-4t^3
}{2}
\Biggr],
\end{split}
\label{eq:thermal_force}
\end{equation}
where $\beta=e^C \approx 1.46$, $\nabla T$ is the local temperature gradient.

\begin{figure}[htbp]
    \centering

    \begin{overpic}[width=0.95\linewidth]{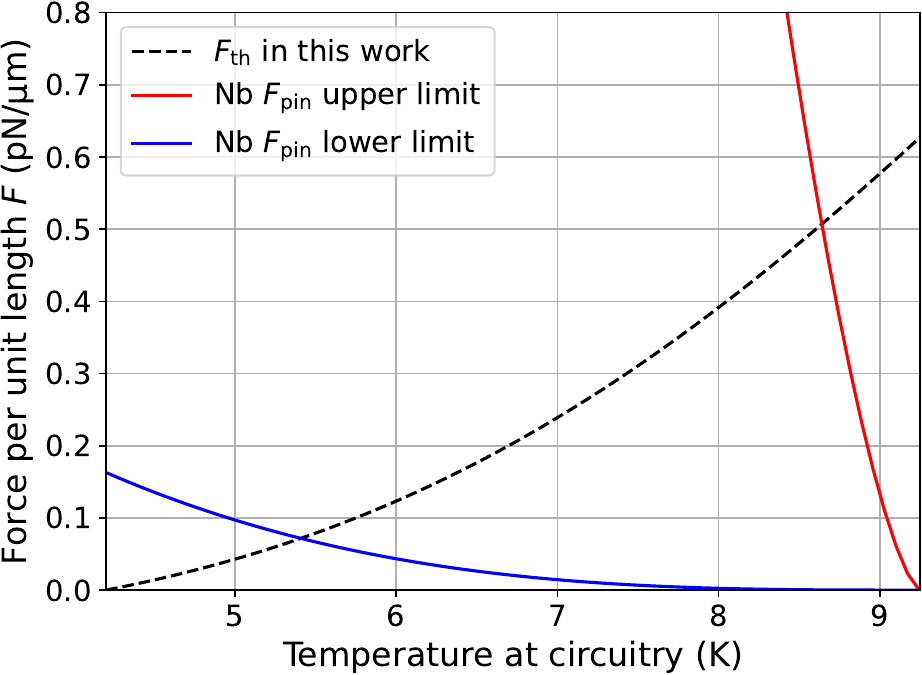}
    \end{overpic}

    \caption{Thermal force, $F_{\mathrm{th}}$, from Eq.~\ref{eq:thermal_force} with $T_c=9.25$ K, $\lambda_0=82$ nm and $\xi_0=18$ nm, at the location of the JJ chain and for the temperature profiles in Fig.~\ref{fig:thermal_modeling_ab}, corresponding to different heat powers $\dot Q$. The pinning force, $F_{\mathrm{pin}}$, is plotted using Eq~\ref{eq:pinning_force} and the reported range of the pinning parameters for Nb. $F_{\mathrm{th}}$ exceeds all reported $F_{\mathrm{pin}}$ values for $T_c-T \geq 500$ mK. In this temperature range, volrtices trapped during field cooling in the JJ chain region, or in other regions of the circuit with even larger temperature gradients and $F_{\mathrm{th}}$, will be driven toward the moats as $F_{\mathrm{th}}$ becomes the dominant force.}
    
    \label{fig:thermal_vs_pinning_force}
\end{figure}

In Fig.~\ref{fig:thermal_vs_pinning_force}, $F_{\mathrm{th}}$ per unit length of the vortex is plotted as a function of local temperature, using the average thermal gradient $\langle\nabla T\rangle$ over the $\pm 1~\upmu\mathrm{m}$ region of the junctions located at the midpoint between the heaters; see Fig.~\ref{fig:thermal_modeling_ab}b. We use $C=0$ to establish a lower limit on $F_{\mathrm{th}}$. As the temperature increases, the average temperature gradient, $\langle \nabla T \rangle$, increases, leading to a larger thermal force $F_{\mathrm{th}}$.

To push vortices into the moats, this thermal force needs to overcome vortex pinning force in Nb films, $F_{\mathrm{pin}}$. Since $F_{\mathrm{pin}}$ can vary widely depending on Nb film preparation, we use the temperature dependence and the range of parameters presented in the literature~\cite{veshchunov2016optical,dew1974flux,dasgupta1978flux,Dhavale_2012,PhysRev.178.657,hovhannisyan2025scanning,horng2002flux,PhysRevB.109.214504,straver2008controlled,PhysRevLett.68.1920}
\begin{equation}
F_{\mathrm{pin}}(T) = F_{\mathrm{pin}}(0)(1-t)^{\gamma}
\label{eq:pinning_force}
\end{equation}
where $F_{\mathrm{pin}}(0)$ is the zero temperature pinning force and $\gamma$ is the scaling factor. Based on the literature, we use $F_{\mathrm{pin}}(0)$ = 30 pN/$\upmu$m and $\gamma$ = 1.5 as an upper limit and $F_{\mathrm{pin}}(0)$ = 1 pN/$\upmu$m and $\gamma$ = 3 as a lower limit. As seen in Fig.~\ref{fig:thermal_vs_pinning_force}, with increasing the local heat power and the corresponding temperature and its gradient, the thermal force at $T_c-T\leq$ 500 mK becomes larger than the pinning force in both limits, indicating that any vortices in the JJ chain or any other region between the heaters will be pushed towards the moats. In the weak pinning regime, this holds true down to $T=5.5$ K. 

As is well known, empty moats attract vortices while moats containing some amount of flux have repulsive potential for vortices at large distances and attractive potential at small distances if the number of flux quanta in the moat is less than the saturation number, $N_s$, at which the moat potential becomes repulsive at any distance~\cite{ginzburg1950zh,ginzburg1958critical,tinkham2004introduction,bardeen1962critical,mkrtchyan1972interaction,kapur2026mitigationmagneticfluxtrapping}. The thermal force appears to not only overcome vortex pinning on defects, but also pushes vortices from the regions of repulsive potential far from the moats into a closer proximity where the moat potential changes to attractive, allowing the moat to absorb additional flux quanta. 

The maximum number of flux quanta that can be stored in a moat, $N_{\mathrm{max}}$, is reached when the screening current on the moat rim reaches the critical current value, representing an upper limit on the extent at which moat aligned resistors can prevent flux trapping. For the square moat with size $a$, $N_{\mathrm{max}}$ can be estimated as $N_{\mathrm{max}}\sim 0.8a/\xi{(T)}$~\cite{ginzburg1950zh,ginzburg1958critical,tinkham2004introduction,bardeen1962critical,mkrtchyan1972interaction,kapur2026mitigationmagneticfluxtrapping}. While the exact temperature at which flux capturing in the moats occurs and thus, the appropriate value for $\xi(T)$ remains unclear, we estimated $N_{\mathrm{max}}$ for $T_c - T \leq 500\,\mathrm{mK}$ using $\xi(T) = \xi_0/{\sqrt{1 - t}}$
with $\xi_0 = 20\,\mathrm{nm}$. For the 10-$\upmu$m square moats used in the circuit, $N_{\mathrm{max}} \sim$ 30 -- 90 flux quanta in the $T_c-T$ range from 50 to 500 mK, respectively. 

If a circuit is built of cells with sizes $p_x\times p_y$ and each cell has a moat with aligned heaters, allowing it  to capture all vortices up to the moat full capacity $N_{\mathrm{max}}$ using the optimized thermal gradient, such a circuit could vortex free in fields up to $B_r=N_{\mathrm{max}} \Phi_0 / (p_x p_y) \sim 150-500\,$ $\upmu$T using $p_x =p_y=20 \, \upmu$m and $a=10 \,\upmu$m square moats. This value could be further improved through optimization of the moat geometry~\cite{kapur2026mitigationmagneticfluxtrapping}. For example, using high-aspect-ratio \(36 \,\upmu \mathrm{m} \times1~\upmu\mathrm{m}\) moats would increase \(N_s\) by approximately a factor of two. However, the practical limit of \(N_{\mathrm{max}}\) at such high fields remains unclear. If the number of vortices pushed into a moat from the film,  \(N_{\mathrm{film}}\), exceeds the moat capacity \(N_{\mathrm{max}}\), the excess vortices will accumulate near the moat but cannot be trapped. As the device cools further, \(N_s\) increases as \(\xi(T)\) decreases, eventually reaching \(N_{\mathrm{max}}>N_{\mathrm{film}}\). It is unclear whether vortices that were not captured into the moat at high local temperatures can subsequently be trapped during the chip cooldown as resistors are turned off, or remain immobilized. Consequently, the relevant value of \(N_{\mathrm{max}}\) may depend not only on the temperature vortices are driven into the moat, but also on the dynamics of vortex transport and moat trapping during cooldown. Experimentally, we observed up to \(16\Phi_0\) trapped within individual moats. This implies \(\xi(T)\lesssim 500~\mathrm{nm}\), corresponding to \(T_c-T\gtrsim 15~\mathrm{mK}\) during flux trapping dynamics.
 
Flux generated by circuit operation or in zero-field-cooled conditions, where the thermal gradients are absent, can likely be mitigated by pulsing the resistors. In this case, the resistors generate sufficiently large thermal gradients to either locally thermal cycle the system or drive vortices away from the circuitry similar to the experiments presented in Fig.~\ref{fig:IC_measurements}b,c.

\section{Conclusion}
Magnetic flux trapping is one of the fundamental phenomena limiting performance and scalability of superconducting integrated circuits. We have demonstrated a circuit defluxing approach based on using local thermal gradients generated by on-chip heaters and moats in superconducting ground planes. Thermal gradients directed toward the moats provide an additional force on vortices in the same direction as the electromagnetic attraction of moats, thereby helping to overcome vortex pinning in the film and sequester flux in the moats away from the flux-sensitive circuitry. Although the induced thermal gradients are relatively small (\(\lesssim 50~\mathrm{mK}/\upmu\mathrm{m}\)), the heaters simultaneously raise the local film temperature, reducing competing vortex pinning forces and allowing the thermal gradient-induced force to dominate. Magnetic imaging and electrical measurements have confirmed removal of magnetic flux trapped in the circuit upon cooldown in ambient magnetic fields \(B_r \leq 60~\upmu\mathrm{T}\). The approach is likely extendable to larger fields. We have also shown that this technique can remove flux generated by large currents during the circuit operation.

While the effectiveness of the moats-plus-thermal-gradient defluxing method was demonstrated on a small part of a 5 mm $\times$ 5 mm chip, this solution can be scaled to an entire circuit. The current MIT LL SFQ5ee fabrication process~\cite{SFQ5ee} contains a dedicated resistor layer used to fabricate the on-chip heaters for this experiment. The design used would take up $\approx 25\%$ of this layer if deployed on the full circuit scale. A more optimized geometry of the moats, e.g., slit-type moats used in~\cite{shregOld,ac_power_sfq,shregNew,kapur2026mitigationmagneticfluxtrapping} with resistors overlapping the slits would likely encompass $\leq 5\%$ of the resistor layer, leaving enough room for the circuitry. 

The presented flux trapping mitigation method is broadly adaptable to any superconducting circuits, as the heaters do not need to share a plane with the circuitry and do not necessarily require a dedicated resistor material because thin superconducting wires driven above their critical current can also be used. A provision for a tunable thermal gradient can also be built into the chip package or chip holder to utilize during the cooldown instead of (or in addition to) being generated on-chip. 
\section{Acknowledgements}
We thank Tom Osadchy for assistance with diamond growth; Peter O'Brien, Jon Wilson, and Matthew Ricci for help with diamond coating; and Tom Grasso, George Haldeman, Christopher Thoummaraj, Ryan Johnson, and Andrew Maccabe for assistance with instrument design, assembly, and testing. We are grateful to Ravi Rastogi and David Kim for overseeing fabrication of the Nb films and test structures used in this work. We thank Joseph Belarge, Danielle Braje, Logan Bishop-Van Horn for helpful discussions. Rohan Kapur thanks John Kim for continued assistance. 

This material is based upon work supported under Air Force Contract No. FA8702-15-D-0001 or FA8702-25-D-B002. Any opinions, findings, conclusions or recommendations expressed in this material are those of the authors and do not necessarily reflect the views of the U.S. Air Force. Notwithstanding any copyright notice, U.S. Government rights in this article are defined by DFARS 252.227-7013 or DFARS 252.227-7014 as detailed above. Use of this article other than as specifically authorized by the U.S. Government may violate any copyrights that exist in this article. The U.S. Government is authorized to reproduce and distribute reprints for Governmental purposes notwithstanding any copyright annotation thereon.
\appendix       
\setcounter{section}{0}
\section{Application to Chip Scale Defluxing}
\label{subsec:chip_scale_circuit_scaling}
The defluxing solution presented in this work based on using local thermal gradients and moats is generalizable to other types of superconducting circuits. While demonstrated here in a multilayer superconducting digital circuit with dedicated moat and resistor layers, as used in standard MIT LL SFQ5ee processes~\cite{SFQ5ee}, the approach is also compatible with simpler fabrication processes containing fewer layers. Even in a single-layer circuit, resistive heaters can be implemented using narrow superconducting wires driven above their critical current into the normal state and creating a thermal gradient towards the chip edges and/or the moats. 

The circuit area occupied by the resistors and the heat power can be reduced by using a different moat and resistor geometry. High aspect ratio rectangular "slit" moats are more space-efficient and improve flux sequestration at the same moat placement pitch~\cite{kapur2026mitigationmagneticfluxtrapping}. Resistors can be run along the slit moats directly underneath or above them in straight lines, reducing the occupied area and increasing the temperature gradient. The circuitry can be shifted away from the symmetrical location between the moats toward the areas of higher thermal gradients. Flux expulsion measurements using circuits designed with moats and heaters in this geometry would be required to further investigate the total power requirements.

\section{Active Microscope Apparatus}
\label{subsec:microscope_active_upgrade}
The cryogenic NV diamond microscope, excluding the circuitry used to apply current to the device under test (DUT), is described in detail in~\cite{qswift_apparatus_paper}. The capability to drive current through the DUT was implemented in this work for the first time, but is presently limited to six electrical feedthroughs due to wiring and thermal constraints.

In this section, we describe the current implementation of the electrical interface as used in these measurements. A fully upgraded version of the apparatus, with expanded electrical connectivity and improved thermal performance, is under development and will be described in a future publication once complete.

The current cryogenic NV diamond microscope apparatus applies current to the DUT via the mounting bracket which contains 4 electrical feedthroughs capable of connecting up to 40 signals (see Fig. 1c in~\cite{qswift_apparatus_paper}). In addition to the four electrical feedthroughs, the upgraded bracket contains a modified MW interposer board and new signal board. The signal board is aligned with the perimeter of the DUT and wirebonded to the DUT to create an electrical connection between the signal board and DUT pads. The electrical signals from the signal board and then wired out of the bracket via the electrical feed throughs. The bracket electrical feedthroughs are connected to the cryostat electrical feedthroughs, where they are then connected to room temperature hardware used to readout/apply current to the electrical lines. The apparatus is currently limited by the number of available electrical lines and their thermalization. Both needs to be improved for imaging an active superconducting circuit typically requiring about 40 signal lines.

\section{Heater Current Polarity Effects}
\label{subsec:DC_expts_discussion}

\begin{figure}[htbp]
    \centering
    \vspace{0.5em}
    \begin{overpic}[width=0.95\linewidth]{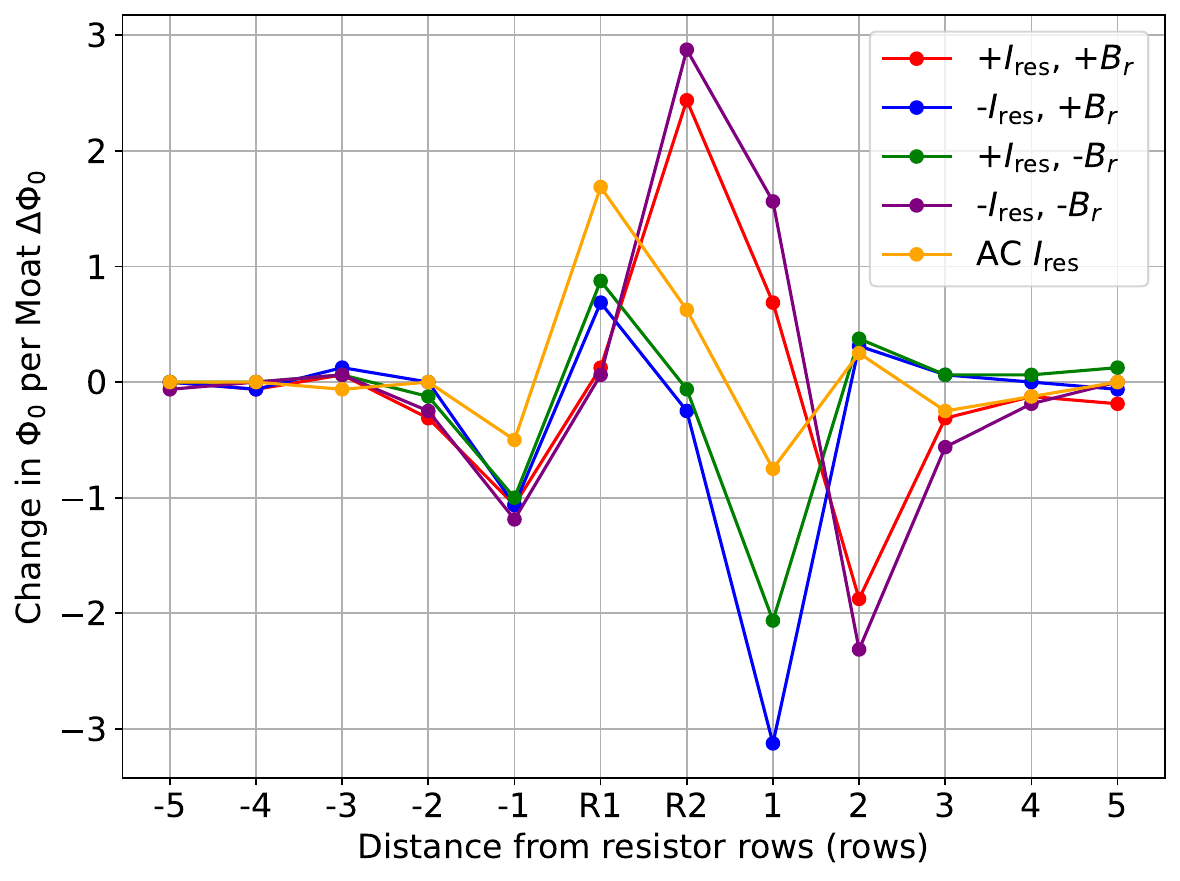}
    \end{overpic}

    \caption{Change in $\Phi_0$ per moat, $\Delta\Phi$, as a function of row position on the chip (centered on the resistor rows), for $I_\mathrm{res}=200~\upmu\mathrm{A}$ and $B_r=31.6~\upmu\mathrm{T}$, with varying relative orientations of $I_\mathrm{res}$ and $B_r$. $\Delta\Phi$ depends strongly on the relative orientation: $(+I_\mathrm{res}, +B_r)$ and $(-I_\mathrm{res}, -B_r)$ produce similar flux distributions, as do $(+I_\mathrm{res}, -B_r)$ and $(-I_\mathrm{res}, +B_r)$, while all DC configurations differ markedly from the AC $I_\mathrm{res}=200$ $\upmu$A case. Since the resistor power is the same for all DC and AC configurations, the flux distribution in the DC configurations is significantly affected by the magnetic field induced by $I_{\mathrm{res}}$, which produces oppositely aligned out of plane magnetic field components on either side of the wire and/or the Lorentz force induced by $I_{\mathrm{res}}$. We defer further discussion of DC $I_{\mathrm{res}}$  effects to Appendix~\ref{subsec:DC_expts_discussion}, however, use an AC $I_\mathrm{res}$ for all other data discussed in this paper as it appears to mitigate the majority of DC related effects.}

    \label{fig:DC_B_field_directionality_vs_delta_flux_quanta}
\end{figure}

We characterized the effects of $I_{\mathrm{res}}$ polarity on the distribution of flux in the chip. We applied a DC current $I_{\mathrm{res}} = 200~\upmu\text{A}$ through both outer moat resistors at $B_r = 31.6~\upmu\mathrm{T}$ as the film was cooled through $T_c$,  and repeated cooldowns using different relative directions of $I_{\mathrm{res}}$ and $B_r$. In Fig.~\ref{fig:DC_B_field_directionality_vs_delta_flux_quanta}, the average change in flux quanta per moat $\Delta\Phi$ is plotted as a function of distance from the resistor rows (in vertical pitch units, corresponding to 20 $\upmu$m).

As seen in Fig.~\ref{fig:DC_B_field_directionality_vs_delta_flux_quanta}, the relative directions of $I_{\mathrm{res}}$ and $B_r$ significantly affect the resulting flux distribution with the $+I_{\mathrm{res}}, +B_r$ and $-I_{\mathrm{res}}, -B_r$ cases yielding similar distributions, while the $-I_{\mathrm{res}}, +B_r$ and $+I_{\mathrm{res}}, -B_r$ cases yield different, but also mutually similar, distributions. For the $+I_{\mathrm{res}}, +B_r$ ($-I_{\mathrm{res}}, -B_r$) case, $\Delta\Phi$ is not consistent across resistor rows with the flux in the R2 resistor row moats significantly increasing, while the flux in the R1 resistor row moats does not increase appreciably. Additionally, the non-resistor moats closest to the R2 resistors also see an appreciable increase in flux. To offset the additional flux in the R2 row, $\Delta\Phi$ is negative for the second nearest neighbors on either side of the R2 resistor row. The flux distribution in the ($-I_{\mathrm{res}}, +B_r$) and ($+I_{\mathrm{res}}, -B_r$) cases is significantly different with $\Delta\Phi$ for R2 being approximately 0 and $\Delta\Phi \approx$ 1 for R1. The moats closest to the resistors see significant decrease in flux quanta to offset this. In all measured field and current distributions, we do not observe a significant change in the total amount of flux quanta in the film per moat $\Delta\Phi_{\mathrm{film}}$ with $\Delta\Phi_{\mathrm{film}}$ for the $+I_{\mathrm{res}}, +B_r$, $-I_{\mathrm{res}}, -B_r$, $-I_{\mathrm{res}}, +B_r$, and $+I_{\mathrm{res}}, -B_r$ cases being -0.04, -0.01, -0.16, and -0.07 $\Phi_0$ per moat. We believe the cause of this slight decrease is flux expulsion to the edges of the film or moats outside the field of view. 

As $I_{\mathrm{res}}$ of both outer moat resistors across all experiments presented in Fig.~\ref{fig:DC_B_field_directionality_vs_delta_flux_quanta} was held constant, the data indicate a strong dependence of the resultant flux distribution on the magnetic fields generated by $I_{\mathrm{res}}$ flowing through the resistors and their orientation relative to $B_r$. This dependence is further confirmed by measurements using AC $I_{\mathrm{res}}$, where the film was cooled with an AC 200 $\upmu$A square wave modulated at 10 Hz applied to both outer moat resistors. As seen in in Fig.~\ref{fig:DC_B_field_directionality_vs_delta_flux_quanta}, changing $I_{\mathrm{res}}$ from DC to AC appeared to suppress the most significant effects arising from the magnetic field of the heaters. In the AC case, the flux in the R1 and R2 moats increases by on average $\approx$ 1 flux quanta and is offset by losses in the moats nearest the resistors, while $\Delta\Phi_{\mathrm{film}}$ is consistent with zero. 
The magnetic field generated by the DC current  produces opposite out of plane field components on opposite sides of the heaters, which adds to the external field $B_r$ and increases the nucleation of vortices on one side of the heater and subtracts on the opposite side, decreasing the concentration of vortices. The perpendicular to the wire palne field component produced by the heater wire a distance $r$ away can be estimated using Ampere's law  $B_{\mathrm{wire}}(r)=\upmu_0 I_{res}/(2\pi r)$, giving $0.2I_{res}/r$ in $\upmu$T with $I_{res}$ in $\upmu A$ and $r$ in $\upmu m$. In the experiments $I_{res}$ was 200 $\upmu$A, producing $B_{\mathrm{wire}}(r)=40/r_{\upmu m}$ $\upmu$T. Since, there are two heaters with the same $I_{res}$ flowing in the same direction, the field in the region between them is reduced, $B(r)=0.40/r_1-40/r_2$, and enhanced in the regions outside $B(r)=40/r_1+40/r_2$, where $r_1$ and $r_2$ are distance from the resistors R1 and R2 to the observation point. This field becomes negligible in comparison to the applied field $B_r$ at distances from the heaters larger than $>40 \, {\mathrm {\upmu m}}$, two rows of moats.

Fig.~\ref{fig:DC_B_field_directionality_vs_delta_flux_quanta} is consistent with additional flux generation of opposite polarities on opposite sides of the wire. In the $+I_{\mathrm{res}}, +B_r$ and $-I_{\mathrm{res}}, -B_r$ cases, the polarity of $B_{\mathrm{wire}}$ is aligned with $B_r$ in rows below the resistors (right part of Fig.~\ref{fig:DC_B_field_directionality_vs_delta_flux_quanta}) and anti-aligned with $B_r$ above the resistors (left part of Fig.~\ref{fig:DC_B_field_directionality_vs_delta_flux_quanta}). As a result, the average field is enhanced in rows R2, 1, 2, almost unchanged between the heaters R1 and R2, and reduced in the rows above R1. Accordingly, we observed additional flux generation in the rows  R2 and 1, while some flux is pulled out of the row 2 by the thermal gradient. In the R1 row, we observed $\Delta{\Phi_{0}} \approx 0$, as additional vortices likely trap in this row of moats from the thermal force, but annihilate with the anti-aligned vortices produced by the R2 resistor field. For the -1 row, $\Delta{\Phi_{0}} < 0$ because the total field in this region is reduced with respect to $B_r$ due to the field produced by both resistors.

In the $+I_{\mathrm{res}}, -B_r$ and $-I_{\mathrm{res}}, +B_r$ cases, the polarity of $B_{\mathrm{wire}}$ is now aligned with $B_r$ above the resistors (left of Fig.~\ref{fig:DC_B_field_directionality_vs_delta_flux_quanta}) and anti-aligned with $B_r$ below the resistors (right of Fig.~\ref{fig:DC_B_field_directionality_vs_delta_flux_quanta}). As a result, the average field distribution inverts with respcet to the described above. Accordingly, we observed $\Delta{\Phi_{0}} \approx 0$ in the R2 row, as additional vortices likely trap in this row from the thermal force, but annihilate with the anti-aligned vortices produced by the R1 resistor row. The 1 row sees $\Delta{\Phi_{0}} < 0$ because the field in this region is reduced with respect to $B_r$. $\Delta{\Phi_{0}} > 0$ in the R1 resistor row due to $B_{\mathrm{wire}}$ of the R2 resistor (in addition to flux added due to the thermal force pulling flux out of row -1).

Interestingly, Fig.~\ref{fig:DC_B_field_directionality_vs_delta_flux_quanta} is not perfectly symmetric about orientations, apparently because the thermal forces are always directed towards the heaters regardless of the current /field directions. In addition, Lorentz force on the vortices, produced by the screening currents in the film,  may also play a role. This asymmetric behavior, however, requires further investigation.

Self-field effects of the heaters described above can be completley eliminated by making each heater of two closely spaced narrow resistive wires with equal currents flowing in opposite directions, e.g., using a narrow u-turn loop. We plan to study this design in combination with slit-type moats in the future work.
\section{Supplementary Images}
\label{subsec:supp_images}
\begin{figure}[htbp]
    \centering
    \vspace{0.5em}
    \begin{overpic}[width=0.95\linewidth,trim=0pt 0pt 40pt 0pt, clip]{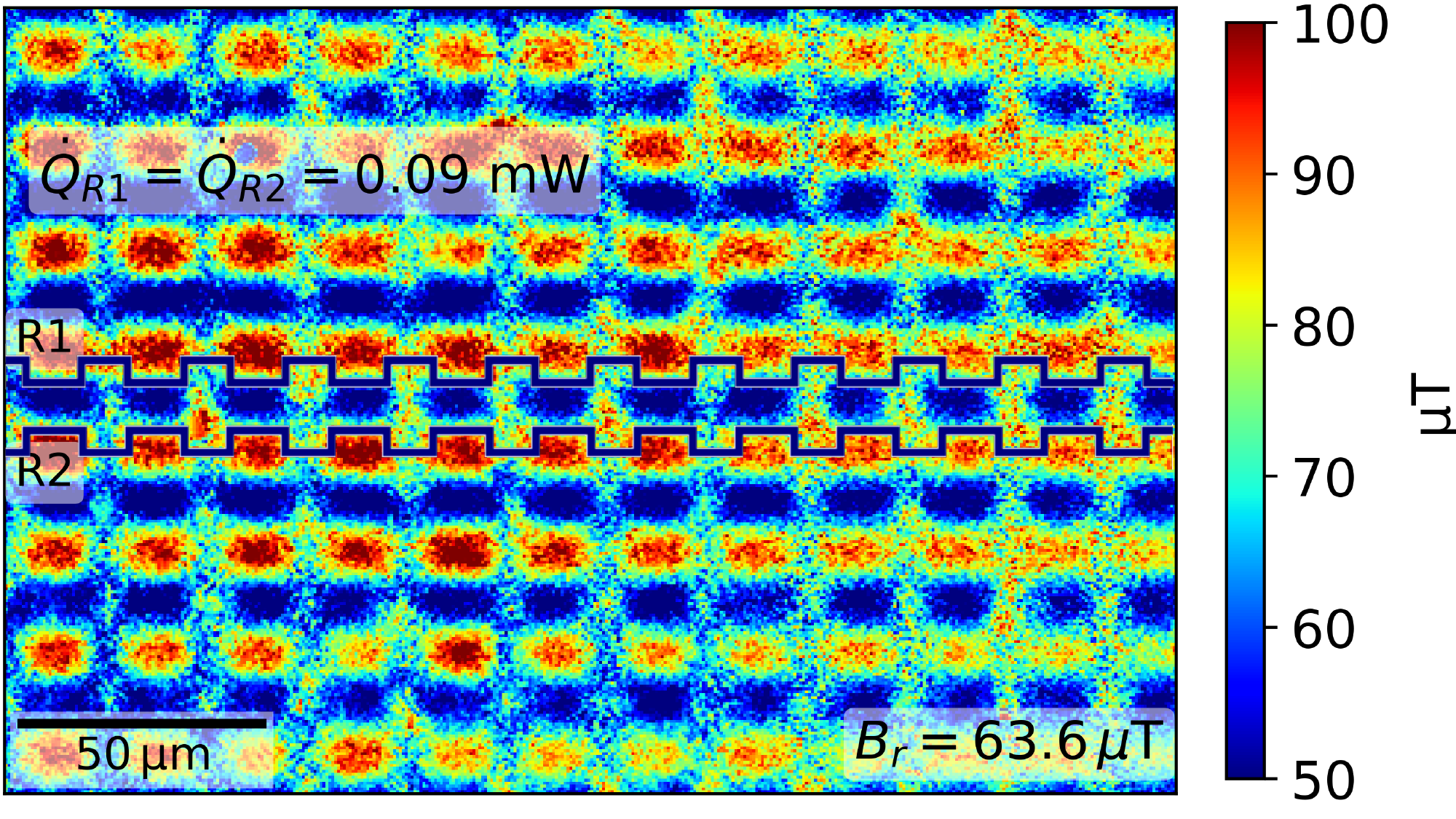}
    \end{overpic}
    \caption{Magnetic image for $B_r = 63.6~\upmu\mathrm{T}$ with $\dot{Q}_{R1}=\dot{Q}_{R2}=0.09$ mW. The outer moat resistors used in the experiments are overlaid in blue in all four images. Compared to Fig.~\ref{fig:thermal_gradient_flux_mitigation}c, where $B_r = 63.6~\upmu\mathrm{T}$ and $\dot{Q}_{R1}=\dot{Q}_{R2}=0$ mW, the vortex distribution in the film outside the moats is very similar, with heating reducing the vortex population in the film near the R1 and R2 moats by approximately $0.1\Phi_0$ per resistor moat.}
    \label{fig:100uA_63_6uT}
\end{figure}
Fig.~\ref{fig:100uA_63_6uT} shows a magnetic image of the circuit for $B_r = 63.6~\upmu\mathrm{T}$ with $\dot{Q}_{R1}=\dot{Q}_{R2}=0.09$ mW. Compared to Fig.~\ref{fig:thermal_gradient_flux_mitigation}c, where $B_r = 63.6~\upmu\mathrm{T}$ and $\dot{Q}_{R1}=\dot{Q}_{R2}=0$ mW, the vortex distribution in the film outside the moats is very similar, with heating reducing the vortex population in the film near the R1 and R2 moats by approximately $0.1\Phi_0$ per moat. In contrast, $\Delta\Phi_0=0.25\Phi_0$ in the R1 and R2 moats and is 0 for all other moats. This indicates that the moats are trapping additional flux across all three ground planes, as the images likely only reflect the vortex population in the top (M7) ground plane.
\begin{figure*}[htbp]
  \centering
  \captionsetup{font=small}
  
  \begin{overpic}[width=0.48\textwidth, trim=140pt 0pt 70pt 0pt, clip]{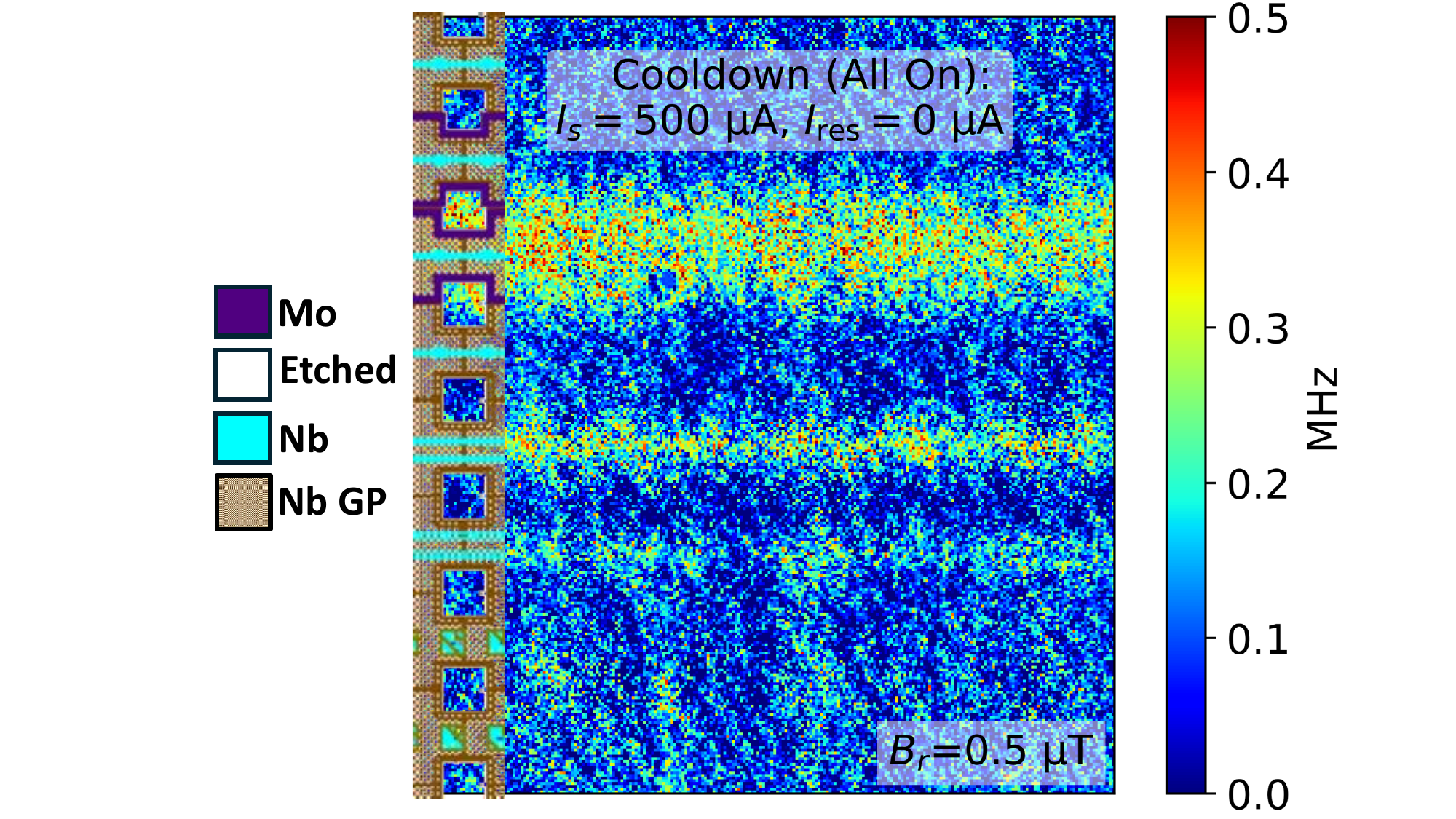}
    \put(-6,52){\footnotesize\textbf{(a)}}
  \end{overpic}
  \hspace{0.02\textwidth}
  \begin{overpic}[width=0.48\textwidth, trim=140pt 0pt 70pt 0pt, clip]{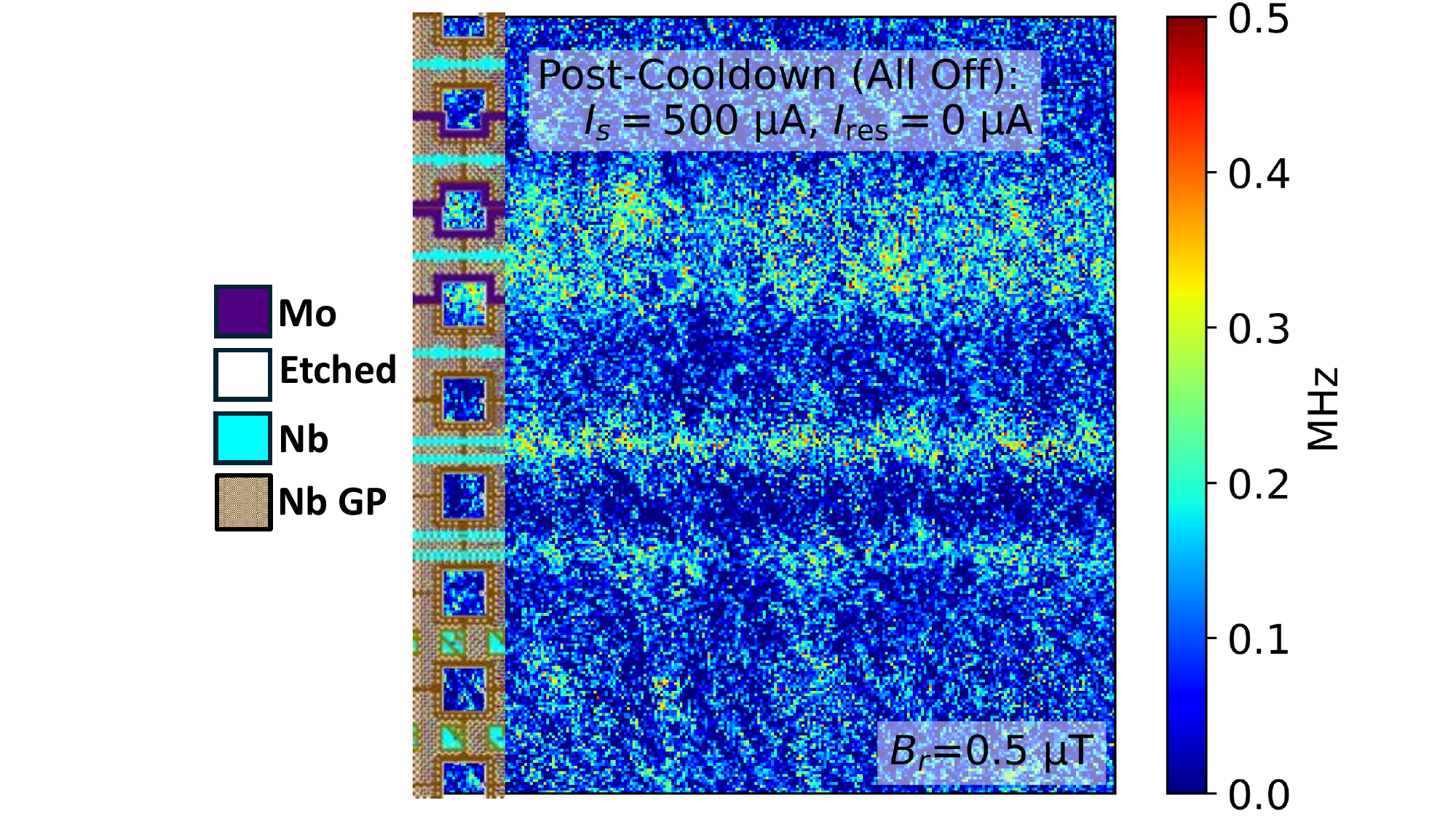}
    \put(-6,52){\footnotesize\textbf{(b)}}
  \end{overpic}

  \vspace{0.75em}

  \begin{overpic}[width=0.48\textwidth, trim=140pt 0pt 70pt 0pt, clip]{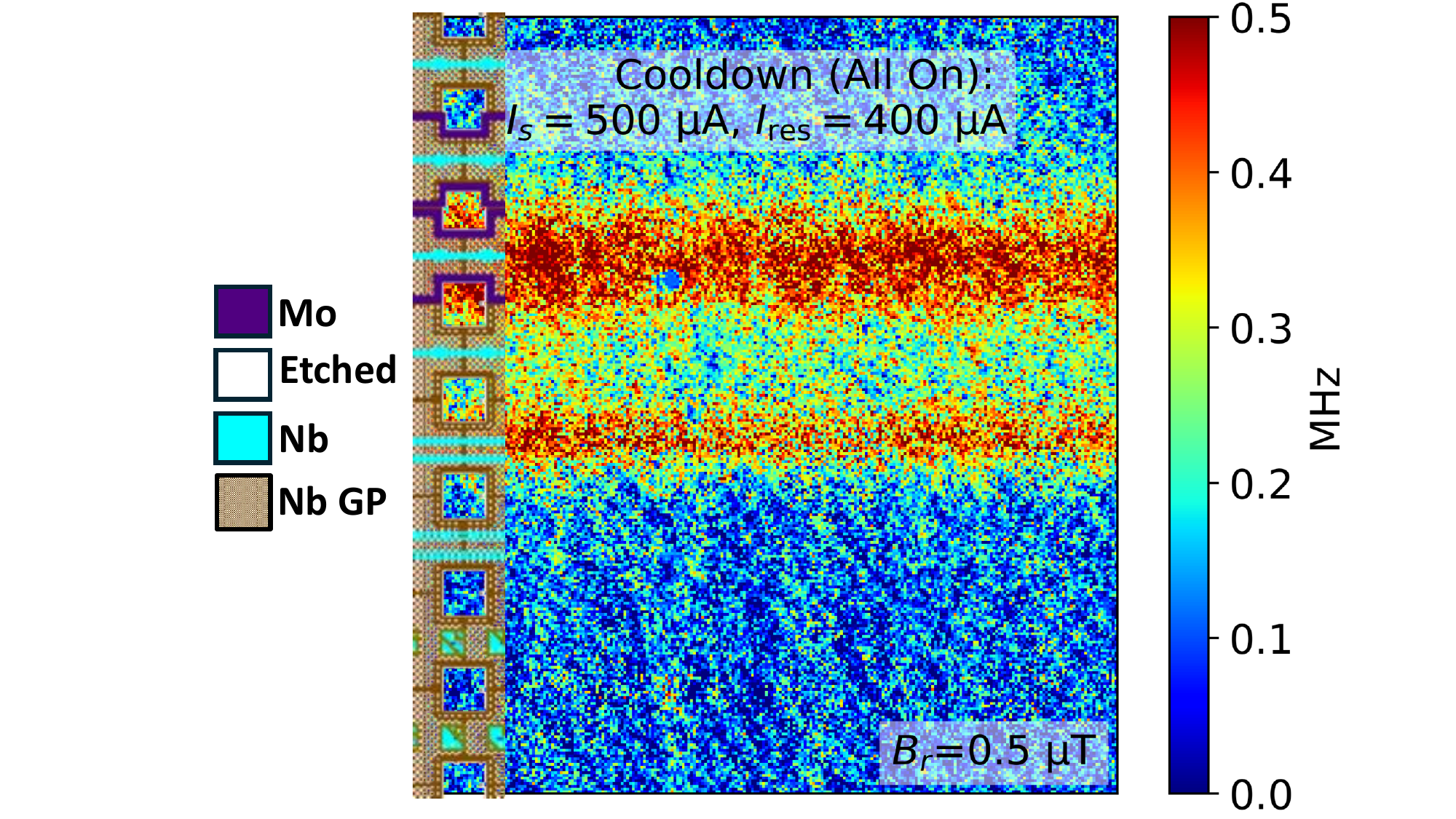}
    \put(-6,52){\footnotesize\textbf{(c)}}
  \end{overpic}
  \hspace{0.02\textwidth}
  \begin{overpic}[width=0.48\textwidth, trim=140pt 0pt 70pt 0pt, clip]{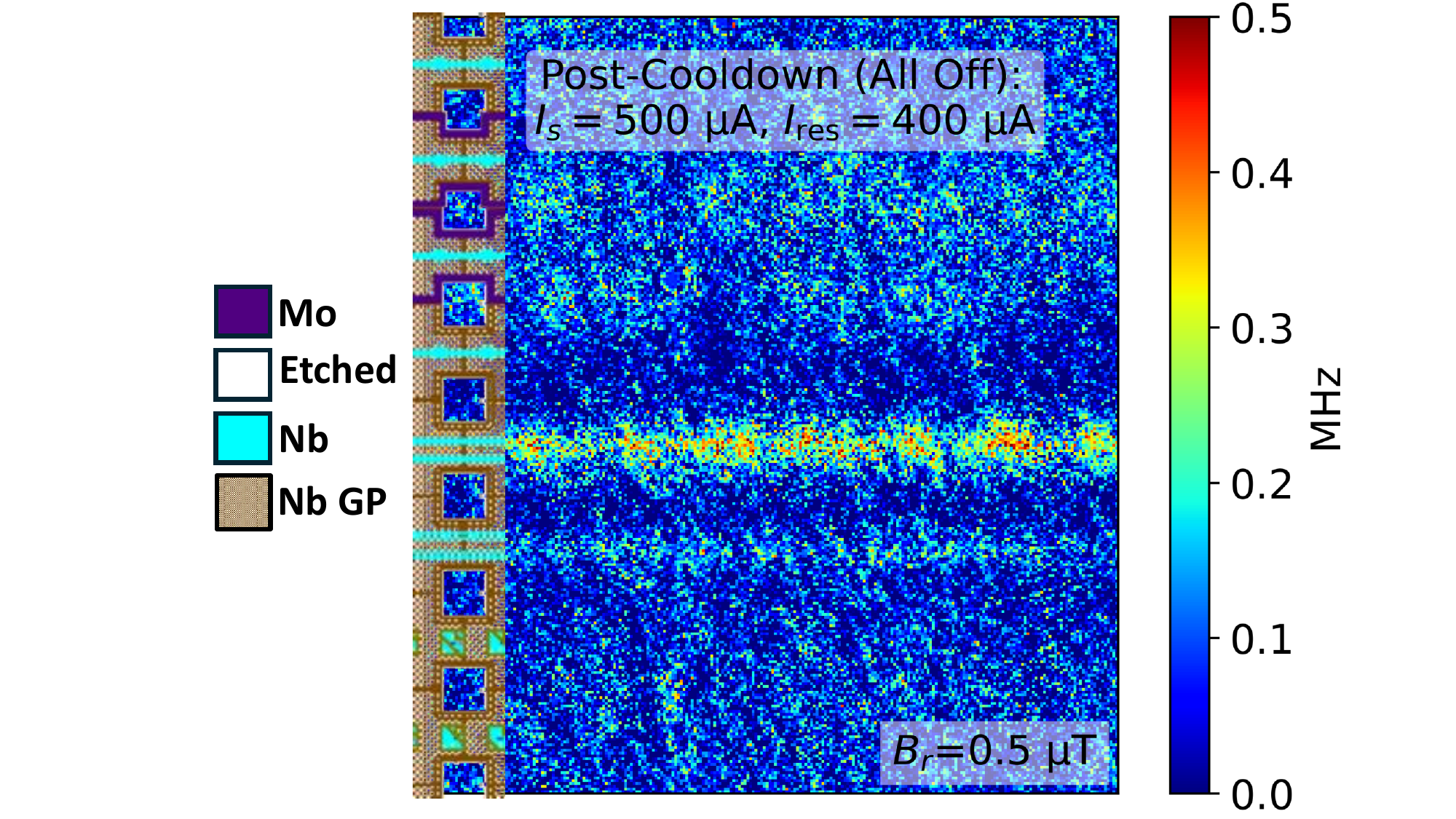}
    \put(-6,52){\footnotesize\textbf{(d)}}
  \end{overpic}

  \caption{(a) Magnetic image of the circuit at base temperature with $I_{s}=500$ $\upmu$A applied to the JJ chain (b) Magnetic image of the circuit with $I_{s}=500$ $\upmu$A applied to the JJ chain during cooldown and shut off once the system reached base temperature (c) Magnetic image of the circuit at base temperature with $I_{s}=500$ $\upmu$A applied to the JJ chain and $I_{res}=400$ $\upmu$A applied to both surrounding outer moat resistors (d) Magnetic image of the circuit with $I_{s}=500$ $\upmu$A applied to the JJ chain and $I_{res}=400$ $\upmu$A applied to both surrounding outer moat resistors during cooldown and shut off once the system reached base temperature. All images were taken at $B_r=0.5$ $\upmu$T. The appearance of enhanced field (vortex density?) between the two Nb wires running parallel to the moats under the top ground plane in all four images corresponds to enhanced trapping of vortices induced by the current through resistors and JJ chains during cooldown. Narrow superconducting wires present a potential barrier for vortices moving in the ground planes toward the moats, associated with vortex entering into the narrow wire and increasing the vortex length and energy. So, some of the vortices become trapped between the two parallel wires despite the thermal gradient forces.
}
  \label{fig:supp_images}
\end{figure*}
Figure~\ref{fig:supp_images} shows the magnetic images corresponding to the JJ suppression and mitigation measurements in Fig.~\ref{fig:IC_measurements}a. Panel (a) shows the system at base temperature with a current $I_s=500\,\upmu\mathrm{A}$ applied to the JJ chain, generating a large local magnetic field in the surrounding region. After the current is turned off, the resulting image in panel (b) reveals significant trapped flux, which suppresses the critical current (see the ``Suppressed'' curve in Fig.~\ref{fig:IC_measurements}a).

Figure~\ref{fig:supp_images}c shows the system again at base temperature, now with current applied to both the JJ chain and the outer moat resistors ($I_s=500\,\upmu\mathrm{A}$ and $I_{\mathrm{res}}=400\,\upmu\mathrm{A}$), modulated as a 10~Hz square wave. Both currents generate strong local magnetic fields, and the thermal gradients produced by the resistors likely help attract flux. After both currents are turned off, the image in panel (d) shows significantly reduced flux near the JJ chain compared to panel (b), with residual flux preferentially located over the outer moat resistors rather than the JJ chain. This redistribution is consistent with the electrical recovery observed in the ``Mitigated'' curve in Fig.~\ref{fig:IC_measurements}a.

\section{Calculation of Thermal Healing Length}
\label{subsec:thermal_healing_length}
The thermal healing length is given by
\[
\lambda_{\mathrm{th}} = \sqrt{\frac{k t}{G}},
\]
where $k$ is the effective in-plane thermal conductivity, $t$ is the effective thickness of the layers participating in lateral heat transport, and $G$ is the thermal conductance per unit area to the bath. The in-plane thermal transport is determined by the two-dimensional (sheet) thermal conductance, $\sum_i f_i k_i t_i$, where $k_i$, $t_i$, and $f_i$ are the thermal conductivity, thickness, and area fraction of layer $i$. In our device, lateral heat flow is dominated by the metallic layers, as Nb and Mo have significantly higher thermal conductivity than the surrounding SiO$_2$ dielectric, whose contribution can be neglected. The superconducting Nb ground planes layers (M1, M4, and M7) have thicknesses of approximately 200~nm, while the Mo heater layer has a thickness of approximately 50~nm and width of only 1 $\upmu$m, so its contribution can be completely neglected. As a result, in-plane heat transport is dominated by the Nb layers, and we approximate the effective thermal conductivity as $k_{\parallel,\mathrm{eff}} \approx k_{\mathrm{Nb}}$.

At low temperatures, thermal conductivity of the Nb films can be estimated using the Wiedemann--Franz law, $k = L_0 T / \rho_0$, where $L_0 = 2.44\times10^{-8}~\mathrm{W\Omega/K^2}$. Using the measured residual resistivity $\rho_0 \approx 3.2~\upmu\Omega\cdot\mathrm{cm}$, we obtain a thermal conductivity near $T_c$, $k_n \approx 7~\mathrm{W,m^{-1}K^{-1}}$. In the superconducting state, electronic thermal transport is reduced due to the opening of the energy gap, leading to an estimated thermal conductivity $k_{\mathrm{Nb}} \sim 1.5~\mathrm{W,m^{-1}K^{-1}}$ at 4.2~K. As the temperature approaches $T_c$, the thermal conductivity approaches the normal-state value.

Because the active circuit layers lie within $\sim 1~\upmu\mathrm{m}$ of the top surface, while the oxidized Si substrate thickness is $\sim 725~\upmu\mathrm{m}$, heat flows predominantly toward the helium bath at the top surface. The heaters are nearly symmetrically located between Nb ground planes M4 and M7 separated by interlayer SiO$*2$ dielectric with thickness $t*{\mathrm{SiO}*2} \approx 1000~\mathrm{nm}$ and thermal conductivity $k*{\mathrm{SiO}*2} \sim 0.1~\mathrm{W,m^{-1}K^{-1}}$~\cite{PhysRevB.4.2029,cahill2003nanoscale}. The thermal resistance between the Nb ground planes, $t*{\mathrm{SiO}*2}/k*{\mathrm{SiO}_2} \approx 10^{-5}~\mathrm{m^2,K,W^{-1}}$, is an order of magnitude smaller than the typical thermal resistance at the SiO$_2$/LHe interface (Kapitza resistance), $R_K \sim 1 \times 10^{-4}~\mathrm{m^2,K,W^{-1}}$ at 4.2~K~\cite{RevModPhys.61.605}.
 Nb layers have negligible contribution to the vertical thermal resistance while all metallic features in the circuit, including wiring, vias, and Josephson junction electrodes, provide conduction pathways that are thermally coupled to the ground planes and can modestly enhance vertical heat spreading between the layers. The top SiO$_2$ passivation layer above M7 is 200\,nm thick and has a thermal resistance of approximately
$2\times10^{-5}\,\mathrm{K\,m^2\,W^{-1}}$. So, the thermal resistance between Nb films and the bath is mainly determined by the thermal resistance $R_K$ of the chip--LHe interface. Therefore, the Nb layers can thermalize before heat is transferred into the LHe bath and be treated as a single composite layer with the total thickness $t \approx 600$~nm. Using the values of $k_n$, $t$, and $R_K$, we obtain $\lambda_{\mathrm{th}} \sim 20~\upmu$m for our system. 
 
Direct measurements of the thermal resistance between the resistor layer and the LHe bath in the SFQ5ee process structure without the M1 and M7 Nb layers gave $R_{\mathrm{bath}} \approx 3\,\mathrm{K\,\upmu m^2/\upmu W}\ \left(3\times10^{-6}\,\mathrm{K\,m^2\,W^{-1}}\right)$~\cite{tolpygo2016superconductor}. Using the latter value of the thermal resistance results in $\lambda_{\mathrm{th}} \sim 3.5\,\upmu\mathrm{m}$. The $\lambda_{\mathrm{th}} = 7.7 \pm 0.5\,\upmu\mathrm{m}$ fitting the data in this work is in between these two estimates.
\section{Circuit Thermal RC Constant}
\label{subsec:circuit_thermal_RC}
The thermal time constant $\tau_{th}=CR_{\mathrm{bath}}$, of heat propagation in Nb films on oxidized Si substrates has been considered multiple times in connection with the development of thermomagnetic instabilitiies and vortex avalanches in Nb films, where $C$ is the effective heat capacitance of the structure per unit area;  see, e.g.~\cite{aranson2005dendritic,baggio2005boundary} and references therein. It is known to be quite short, in the $10^{-6}-10^{-7}$ s range, for films with $t \sim 0.5\, \upmu$m; see~\cite{abaloszewa2026thermally} and references therein. The low limit can be estimated if we assume that the entire 725-$\upmu$m-thick Si substrate contributes to the total heat capacitance of the structure. Its contribution at 9 K can be estimated as $3.4\times10^{-1}~\mathrm{J\,m^{-2}K^{-1}}$, resulting in $\tau_{th} \approx 34 ~\mathrm{\upmu}$s.
This short temperature relaxation time justifies the use of the stationary heat diffusion equation in the main text.

\bibliography{biblio.bib}

\end{document}